\documentclass[12pt,a4paper]{article}  
\usepackage{amsmath}
\usepackage{amssymb}
\usepackage{epsfig}
\usepackage{epstopdf}
\usepackage{graphicx}
\usepackage{subfigure}
\usepackage{commath}
\usepackage{float}
\usepackage[left=2.0cm,right=2.0cm, top=2.5cm,bottom=2.5cm]{geometry}
\usepackage{cite}
\usepackage{times}  
\usepackage[colorlinks=true, linkcolor=blue, citecolor=red, urlcolor=blue]{hyperref} 

\begin{document} 
\newcommand{\sheptitle}
{Deviation from $\mu-\tau$ reflection symmetry under radiative corrections in the minimal seesaw framework}
\newcommand{\shepauthor}
{ Prokash Pegu\footnote{ E-mail: peguprokash202@gmail.com} and Chandan Duarah \footnote{ E-mail: chandanduarah@dibru.ac.in}}
\newcommand{\shepaddress}
   { Department of Physics, Dibrugarh University,
               Dibrugarh - 786004, India }
\newcommand{\shepabstract} 
{The $\mu$-$\tau$ reflection symmetry predicts a maximal atmospheric mixing angle, $\theta_{23}=\pi/4$, and Dirac CP phase, $\delta=\pi/2$ or $3\pi/2$. Recent global analyses indicate small but significant deviations from these predictions, suggesting that the symmetry is approximate and requires breaking. Motivated by this, we study the breaking of $\mu$-$\tau$ reflection symmetry induced by radiative corrections in the minimal seesaw framework, assuming the symmetry to be exact at the high-energy seesaw scale $\Lambda_{\mu\tau}=10^{14}~\mathrm{GeV}$. A distinctive feature of the minimal seesaw is that one light neutrino remains massless, leaving only one physical Majorana CP phase. Starting from the integral solution of the one-loop RGE for the effective Majorana neutrino mass matrix, we derive analytical expressions for the low-energy neutrino parameters at $\Lambda_{\mathrm{EW}}=172.76~\mathrm{GeV}$ in terms of their high-energy counterparts. We then numerically estimate the low-energy parameters within the MSSM, taking $\Lambda_S=1~\mathrm{TeV}$ and $\tan\beta=10$, $30$ and $50$. The analysis is performed separately for Normal Order (NO) and Inverted Order (IO). We find that the predicted neutrino masses and mixing parameters are consistent with current experimental data in both scenarios. The sum of neutrino masses, Jarlskog invariant $J$, and effective Majorana mass $|\langle m\rangle_{ee}|$ also satisfy current experimental constraints. Finally, we estimate the amount of deviations of the low-energy neutrino parameters from their high-energy values and investigate their dependence on $\tan\beta$. We find that the magnitude of these deviations increases with increasing $\tan\beta$ for both NO and IO.\\
Key-words: Lepton mixing, $\mu-\tau$ reflection symmetry, minimal seesaw framework, radiative corrections, MSSM.
 }
\begin{titlepage}
\begin{flushright}
\end{flushright}
\begin{center}
{\large{\bf\sheptitle}}
\bigskip\\
\shepauthor
\\
\mbox{}\\
{\it\shepaddress}\\
\vspace{.5in}
{\bf Abstract }
\bigskip
\end{center}
\setcounter{page}{0}
\shepabstract
\end{titlepage}

\section{Introduction}
\indent~ Evidence from neutrino oscillations experiments has shown that neutrinos are massive particles and that the flavor eigenstates are mixtures of the mass eigenstates. The phenomenon of neutrino oscillation is theoretically explained by the framework of neutrino mixing. Over the past two decades, solar, atmospheric, reactor, and accelerator neutrino oscillation experiments have significantly improved our understanding of neutrino mixing parameters. As a result, some neutrino mixing parameters have now been measured with remarkable precision. Neutrino oscillation experiments have provided precise values of the solar mixing angle $\theta_{12}$, reactor mixing angle $\theta_{13}$ along with the two mass-squared differences $\Delta m^2_{21}=m^2_2-m^2_1$ and 
 $\vert{\Delta m^2_{31}}\vert=\vert{m^2_3-m^2_{1}}\vert/  \vert{\Delta m^2_{32}} \vert =\vert{m^2_3-m^2_{2}}\vert$ \cite{SK2,SNO1,MINOS1,DAYA1,Reno,Dchooz,T2K1,Nova2,ICE3,global2}. In addition, these experiments indicate that the atmospheric mixing angle $\theta_{23}$ is close to $45^\circ$, suggesting nearly maximal atmospheric mixing. However, the exact value of the atmospheric mixing angle $\theta_{23}$ remains uncertain. In particular, it is still not known whether $\theta_{23}$ lies in the first octant ($\theta_{23}<45^\circ$) or the second octant ($\theta_{23}>45^\circ$), a problem commonly referred to as the octant degeneracy. Further, another important unresolved issue is the neutrino mass order, namely whether the neutrino masses follow the normal order ($m_1<m_2<m_3$) or the inverted order ($m_3<m_1<m_2$). The best-fit values and $3\sigma$ ranges of mixing parameters are presented in Table 1 \cite{global2}. Note that the oscillation experiments are insensitive to the absolute values of the mass eigenvalues. We obtain an upper bound for the sum of mass eigenvalues $\Sigma \vert{m_i}\vert < 0.12\ eV$ ($i=1, 2, 3$) from cosmological observations \cite{Mass1}. In addition, neutrino oscillation experiments are also insensitive to Majorana CP phases $\rho$ and $\sigma$. The only experimental way to probe these phases is through neutrinoless double beta decay ($0\nu\beta\beta)$. The decay width of $0\nu\beta\beta$ contains the effective Majorana neutrino mass  $|\langle m \rangle_{ee}|$. The most stringent upper bound on $|\langle m \rangle_{ee}|$ is currently provided by the KamLAND-Zen Collaboration, which reports  $|\langle m \rangle_{ee}|<(0.028-0.122)\ eV$ \cite{Kam}.\\
\indent~ Current results from T$2$K, NO$\nu$A and global fits show that the best-fit value of the Dirac CP phase $\delta$ is $274^\circ$ for IO \cite{T2K1,Nova2,global2}. This indicates that the experimentally measured value of $\delta$ is attractively close to the maximal value of $270^\circ$. In view of this, the role of $\mu-\tau$ reflection symmetry becomes more prominent in explaining the lepton flavour mixing pattern, as it accommodates both maximal CP violation ($\delta= (\pi/2)/(3\pi/2)$) and maximal atmospheric mixing ($\theta_{23}=\pi/4$) along with non-zero $\theta_{13}$. The concept of this symmetry was first put forwarded by Harrison and Scott \cite{ref1}. They proposed a specific form of the lepton mixing matrix, which remains invariant under a combined operation of interchanging the $\mu$ and $\tau$ rows and taking the complex conjugate of the entire matrix. This symmetry operation is called the $\mu$-$\tau$ reflection operation and the symmetry of the mixing matrix is called the $\mu$-$\tau$ reflection symmetry. In mass matrix context, this symmetry can be realized in the neutrino mass term, which remains invariant under the $\mu$-$\tau$ reflection operation. A detailed review of this symmetry can be found in Ref.\cite{rev1}. The phenomenological implications of $\mu$-$\tau$ reflection symmetry in different theories of massive and mixed neutrinos have been studied in the literature \cite{vien1, vien2, vien3, king5, zhao, xinggg, nnath1, nishi, nishi1, dirac1,zhou}. Relevant studies also suggest that the implementation of $\mu$-$\tau$ reflection symmetry in the Type-I seesaw framework provides an attractive and well-motivated approach to understanding the observed pattern of neutrino masses and mixing. Note that the Type-I seesaw mechanism is a well-established framework for generating tiny neutrino masses, in which three heavy right-handed neutrino fields  $N_R$ are introduced into the Standard Model (SM). After integrating out these right-handed neutrino fields, an effective Majorana mass matrix for the light neutrinos arises. However, despite its merits, the Type-I seesaw mechanism involves many free parameters, which substantially reduces its predictive power. To improve its predictability, various approaches have been proposed to reduce the number of free parameters. One simple approach is to consider a minimal version, known as the minimal seesaw framework, which involves only two right-handed neutrinos \cite{original}. The significant feature of this minimal seesaw is that one of the three light neutrino masses must be zero. Relevant studies on minimal seesaw can be found in Refs .~\cite {mini1,mini2,mini3,mini4,prog}. \\
\begin{table}[t]
\centering 
\begin{tabular}{l l l l l}
\hline
& Without SK\\
\hline
&Normal Order & & Inverted Order \\
 \hline
Parameter & Best-fit & $3\sigma$ & Best-fit & $3\sigma$\\
\hline
$\sin^2\theta_{12}$ & 0.307 & 0.275-0.345 & 0.308 & 0.275-0.345\\
$\sin^2\theta_{23}$ & 0.561 & 0.430-0.596 & 0.562 & 0.437-0.597\\
$\sin^2\theta_{13}$ & 0.02195 & 0.02023-0.02376 & 0.02224 & 0.02053-0.02397\\
$\delta$ & 177 & 96-422&  285 & 201-348\\
$\Delta m^2_{21} (\setminus10^{-5} e{V^2})$ & 7.49 & 6.92-8.05& 7.49& 6.92-8.05\\
$\Delta m^2_{32 (31)} (\setminus10^{-3} e{V^2})$ & 2.534 & 2.463-2.606& -2.510 & -2.584- (-2.438)\\
\hline
& With SK \\
\hline
Parameter & Best-fit & $3\sigma$ & Best-fit & $3\sigma$\\
\hline
$\sin^2\theta_{12}$ & 0.308 & 0.275-0.345 & 0.308 & 0.275-0.345\\
$\sin^2\theta_{23}$ & 0.470 & 0.435-0.586 & 0.550 & 0.440-0.584\\
$\sin^2\theta_{13}$ & 0.02215 & 0.02030-0.02388  & 0.02231 & 0.02060-0.02409\\
$\delta$ & 212 & 214-364&  274 & 201-335 \\
$\Delta m^2_{21} (\setminus10^{-5} e{V^2})$ & 7.49 & 6.92-8.05& 7.49& 6.92-8.05\\
$\Delta m^2_{32 (31)} (\setminus10^{-3} e{V^2})$ & 2.513 & 2.451-2.578& -2.484 & -2.547-(-2.421)\\
\hline
\end{tabular}
\caption{Best-fit values and $3\sigma$ allowed ranges of the neutrino oscillation parameters for NO and IO as per global analysis \cite{global2}.}
\end{table}
\indent~  In general, it is hard to believe the exact maximality of the atmospheric mixing angle and$/$or CP violation as constrained by the $\mu$-$\tau$ reflection symmetry. It therefore becomes necessary to study the possible perturbation to $\mu$-$\tau$ reflection symmetry. In this work, we consider breaking of $\mu-\tau$ reflection symmetry  as a result of radiative corrections in the minimal seesaw framework. The $\mu$–$\tau$ reflection symmetry is assumed to be preserved at a flavour symmetry scale usually chosen at the seesaw scale and consider its breaking due to the RG running from the flavour symmetry scale to the low energy scale. As the energy scale decreases, the physical parameters, namely the neutrino mass eigenvalues, mixing angles and CP phases, evolve according to their Renormalization Group Equations (RGEs) and consequently the symmetry breaks down. We employ the run-and-diagonalize RG approach to study the breaking of the $\mu$–$\tau$ reflection symmetry. In this approach, the breaking of symmetry is taken account through the integral solution of the relevant RGE of the effective Majorana neutrino mass matrix $M_\nu$. Adopting the integral solution method to the one-loop RGE of $M_\nu$, we derive analytical expressions for all the physical parameters- masses, mixing angles and CP phases. These expression enables us to express all the low energy parameters in terms of the corresponding high energy parameters. Further, these  also allow us to study correlations between the high-energy parameters and the low-energy parameters. Using these expressions, we then study the numerical predictions of all parameters at the low energy scale $\Lambda_{EW}$ by using input values of the parameters at the flavor symmetry scale $\Lambda_{\mu\tau}$. We consider the low energy scale at the top-quark mass scale, $\Lambda_{\text{EW}} = m_t = 172.76\ \text{GeV}$  and  the flavor symmetry scale at the seesaw scale, $\Lambda_{\mu\tau} = 10^{14}\text{GeV}$. The analysis is done in the MSSM framework and we consider a SUSY-breaking scale at $\Lambda_s= 1 \ TeV$, intermediate between $\Lambda_{\mu\tau}$ and $\Lambda_{EW}$. Since the exact value of the SUSY-breaking scale is not yet known, the choice of $\Lambda_s=1\ TeV$ is motivated by recent studies \cite{Ssy13,Ssy14,Ssy15}. As the reflection symmetry is considered to be preserved at the seesaw scale, the input values of some parameters (atmospheric mixing angle, CP phases) are taken as constrained by the symmetry while input values of the remaining parameters, viz.,  mass eigenvalues, reactor mixing angles, solar mixing angles are estimated from correlation studies. With appropriate input values chosen from correlation studies at $\Lambda_{\mu\tau}$, we first obtain the values of all parameters at $\Lambda_s$, and then use them to obtain the corresponding low-energy predictions at $\Lambda_{EW}$. The present framework allows us to  derive all analytical expressions for the mixing parameters separately for the Normal Order and Inverted Order scenarios. Accordingly, the numerical predictions are also obtained separately for each scenario.\\
\indent~ The purpose of this work is to study the  deviations from $\mu$-$\tau$ reflection symmetry as a consequence of radiative corrections in the minimal seesaw framework. Similar kinds of studies have also been made in some previous works  \cite{nnath,liu}. However, present work differs from the earlier works in certain aspects. First, we have derived analytical expressions for the neutrino parameters, those establish a direct connections between the low energy parameters and high energy parameters. Each of the low energy parameter is completely expressible in terms of corresponding high energy parameters. Second, while numerically estimating the amount of deviations from the $\mu$-$\tau$ reflection symmetry, we consider a SUSY breaking scale at $1\ TeV$ which is intermediate between the flavor symmetry scale $(\Lambda_{\mu\tau}=10^{14}\ GeV)$ and the low energy scale $(\Lambda_{EW}=172.76\ GeV)$. This differs from the situation where the SUSY breaking scale is often taken at the electroweak scale itself. Last but not the least, we adopt a specific assignment of maximal values of CP phases in realizing the $\mu$-$\tau$ reflection symmetry in the lepton mixing matrix. Since $\mu$-$\tau$ reflection symmetry is a form of generalized CP symmetry, there, in general, exist several possible ways of choosing the maximal values of CP phases. The particular maximal assignments of CP phases adopted in this work maintain the reflection symmetry in the standard parametrization, which is consistent with the Harrison-Scott formulation of $\mu$-$\tau$ reflection symmetry.\\
\indent~ The rest of this paper is organized as follows: in section 2, we present some basics of $\mu$-$\tau$ reflection symmetry and the parameterization of the lepton mixing matrix. In section 3, we briefly review the realization of the $\mu$-$\tau$ reflection symmetry in the minimal seesaw framework. In section 4, we present the analytical treatment of radiative corrections to $\mu$-$\tau$ reflection symmetry in the minimal seesaw framework. In section 5, we perform a numerical analysis to estimate the low-energy values of the neutrino parameters using the high-energy parameters as input. We also estimate the amount of deviation of the low-energy neutrino parameters from their corresponding high-energy input values and see how the predicted low-energy parameters vary with $\tan\beta$.  Finally, section 6 is devoted to a summary and discussion.
\section{Some basics of $\mu$-$\tau$ reflection symmetry and parameterization of lepton mixing }
\indent~ The $\mu$-$\tau$ reflection symmetry may be defined as a symmetry of the neutrino mass term under the transformations \cite{rev1} : 
\begin{equation}
\nu_e \rightarrow \nu^c_{eL}, \quad \nu_{\mu} \rightarrow \nu^c_{\tau L}, \quad \nu_{\tau} \rightarrow \nu^c_{\mu L},\label{eq:1}
\end{equation}
where $\nu_{eL}$, $\nu_{\mu L}$, and $\nu_{\tau L}$ denote the left-handed neutrino fields and the superscript $'c'$ represents charge conjugation. In the flavor basis, the effective Majorana neutrino mass term that remains invariant under the above transformations can be written as 
\begin{equation}
\mathcal{L_\nu}=\overline{\left(\begin{array}{ccc}
\nu_e &\nu_\mu &\nu_\tau 
\end{array}\right)_L} M_\nu \left(\begin{array}{c}
\nu_e^c\\

\nu_\mu^c\\
\nu_\tau^c
\end{array}\right)_R, \label{eq:2} \end{equation}
where the subscripts $L$ and $R$ denote the left-handed and right-handed neutrino flavor eigenstates respectively and $M_\nu$
is the effective Majorana neutrino mass matrix. The most general form of $M_\nu$ is 
\begin{equation}
M_\nu = \left(\begin{array}{ccc}
M_{ee} & M_{e\mu} & M_{e\tau}\\
M_{e\mu } & M_{\mu \mu} & M_{\mu\tau}\\
M_{e\tau } & M_{\mu\tau } & M_{\tau\tau}
\end{array}\right),\label{eq:3}
\end{equation}
where all the entries are, in general, complex. Imposing the transformation given in Eq.(\ref{eq:1}) on the mass term in Eq.(\ref{eq:2}) leads to a specific form of the Majorana neutrino mass matrix, whose elements satisfy the relations
\begin{equation}
    \begin{aligned}
        M_{ee}=M_{ee}^*,\ M_{e\mu}=M_{e\tau}^*,\ M_{\mu \mu}= M_{\tau\tau}^*,\ M_{\mu\tau}=M_{\mu\tau}^*.\label{eq:4}
    \end{aligned}
\end{equation}
A mass matrix obeying these relations is said to possess $\mu$-$\tau$ reflection symmetry. We denote the corresponding symmetric mass matrix by $M_\nu^{\mu\tau}$ such that
\begin{equation}
M_\nu^{\mu\tau} = \left(\begin{array}{ccc}
m_{ee} & m_{e\mu} &  m_{e\mu}^*\\
m_{e\mu} & m_{\mu \mu} & m_{\mu\tau}\\
 m_{e\mu}^* & m_{\mu\tau} &  m_{\mu \mu}^*
\end{array}\right).\label{eq:5}
\end{equation}
Note that the elements $m_{ee}$ and $m_{\mu\tau}$ in the above mass matrix become real as a consequence of the reflection symmetry, while other elements remain complex.
 The symmetry in the above mass matrix can also be realized in terms of a $\mu$-$\tau$ exchange operator, given by
\begin{equation}
 A_{\mu \tau} =\left(\begin{array}{ccc}
1&0&0\\
0&0&1\\
0&1&0
\end{array}\right),\label{eq:6}
\end{equation}
such that 
\begin{equation}
(A_{\mu\tau}M_\nu^{\mu\tau}A_{\mu\tau})^*= M_\nu^{\mu\tau}\label{eq:7}
.\end{equation} 
\indent~The effective Majorana neutrino mass matrix in Eq.(\ref{eq:3}) can be diagonalized by the lepton mixing matrix $U$ as
\begin{equation}
U^{\dagger}M_{\nu}U^{*}
= D_{\nu}
= \left(\begin{array}{ccc}
m_1&0&0\\
0&m_2&0\\
0&0&m_3
\end{array}\right),\label{eq:8}
\end{equation}
where $m_{1}$, $m_{2}$, and $m_{3}$ denote the neutrino mass eigenvalues. The lepton mixing matrix is in general expressed as $U = U_{\ell}^{\dagger}U_{\nu}$ ,
where $U_{\ell}$ and $U_{\nu}$ are the unitary matrices that diagonalize the charged-lepton mass matrix and left-handed Majorana neutrino mass matrix respectively. In the basis where the charged-lepton mass matrix is diagonal ( $U_{\ell}$ becomes an identity matrix), we have $U = U_{\nu}$. In general, the parametrization of the lepton mixing matrix can be expressed as
\begin{equation}
  U =P_1 V P_2,\label{eq:9}
\end{equation}
where $P_1=Diag \left( e^{i \phi_1} , e^{i \phi_2}, e^{i \phi_3}\right)$ and $P_2=Diag\left(e^{i\rho}, e^{i\sigma},1\right)$ are two diagonal matrices which contain the three un-physical phases $\phi_1$, $\phi_2$, $\phi_3$ and two Majorana phases $\rho$ and $\sigma$ respectively. The matrix $V$ is given by
\begin{equation}
V = \left(\begin{array}{ccc}
c_{12}c_{13}&s_{12}c_{13}&s_{13}{e}^{-i\delta}\\
-s_{12}c_{23}-c_{12}s_{23}s_{13}{e}^{i\delta}&c_{12}c_{23}-s_{12}s_{23}s_{13}{e}^{i\delta}&s_{23}c_{13}\\
s_{12}s_{23}-c_{12}c_{23}s_{13}{e}^{i\delta}&-c_{12}s_{23}-s_{12}c_{23}s_{13}{e}^{i\delta}&c_{23}c_{13}
\end{array}\right),\label{eq:10}
\end{equation}
where $s_{ij}=\sin\theta_{ij}$ and $c_{ij}=\cos\theta_{ij }$ with $ij=12, 23, 13$.\\
\indent~ The diagonalization of the reflection symmetric mass matrix $M_\nu^{\mu\tau}$ in Eq.~(\ref{eq:5}) requires a mixing matrix that respects the same symmetry. Mathematically, the $\mu$-$\tau$ reflection symmetry in the lepton mixing matrix can be expressed as \cite{prog}
\begin{equation}
U=A_{\mu\tau}U^{*},
\label{eq:11}
\end{equation}
where $A_{\mu\tau}$ is defined in Eq.~(\ref{eq:6}). The right-hand side of the above equation expresses the action of the $\mu$-$\tau$ reflection operation on $U$, $U^*$ being the complex conjugate of $U$. Given the parameterization of $U$ in Eq.(\ref{eq:9}), the invariance condition in Eq.(\ref{eq:11}) leads to the well known maximal predictions:
\begin{equation}
\theta_{23}=\frac{\pi}{4}, \qquad
\delta=\frac{\pi}{2}\ \text{or}\ \frac{3\pi}{2}.\label{eq:12}
\end{equation}
The invariance condition of $U$ also puts constraints on the Majorana phases and the unphysical phases to be either $0$ or $n\pi/2$ $(n=1,\ 2...)$. \\
\indent~ As per the constraints of  $\mu$-$\tau$ reflection symmetry, we choose, in this work, two specific sets of values for the Majorana phases and unphysical phases as considered in Ref.\cite{duarah}. This enables us to maintain the reflection symmetry in the standard parametrization of the lepton mixing matrix in consistent with the original formulation by Harrison and Scott (HS) \cite{ref1}. In their work, Harrison and Scott formulated a specific parametrization of the lepton mixing matrix given by 
\begin{equation}
U_{HS}=\left(\begin{array}{ccc}
u_1&u_2&u_3\\
v_1&v_2&v_3\\
v_1^*&v_2^*&v_3^*
\end{array}\right)
,\label{eq:13}\end{equation}
where first row elements ${u_i}$'s are restricted to be real and ${v_i}$'s are complex. It is easy to see that the real nature of ${u_i}$'s is necessary to satisfy the invariance condition Eq.~(\ref{eq:11}). To maintain consistency with HS formulation of reflection symmetry, we have chosen the allowed values of Majorana phases and unphysical phases such that the first row elements of $U$ become real. Before presenting these choices, we first identify two separate cases based on the values of the Dirac CP phase allowed by the $\mu$–$\tau$ reflection symmetry:
\begin{itemize}
\item \textbf{Case-I:} \ $ \theta_{23}^{\mu\tau}= \pi/4$ and $\delta^{\mu\tau}= \pi/2$,
\end{itemize}
\begin{itemize}
\item \textbf{Case-II:}\ $\theta_{23}^{\mu\tau}= \pi/4$ and $\delta^{\mu\tau}= 3\pi/2$.
\end{itemize}
We then choose the values of Majorana and unphysical phases for the above two cases as
\begin{equation}
\rho^{\mu\tau} = \sigma^{\mu\tau} = \frac{3\pi}{2},
\qquad
\phi_1^{\mu\tau} = \frac{\pi}{2},
\qquad
\phi_2^{\mu\tau} = \phi_3^{\mu\tau} = 0
\label{eq:14}
\end{equation}
for Case-I and 
\begin{equation}
\rho^{\mu\tau} = \sigma^{\mu\tau} = \frac{\pi}{2},
\qquad
\phi_1^{\mu\tau} = \frac{3\pi}{2},
\qquad
\phi_2^{\mu\tau} = \phi_3^{\mu\tau} = 0
\label{eq:15}
\end{equation}
for Case-II. With these choices, we get the lepton mixing matrix from Eq.~(\ref{eq:9}) that possesses $\mu$-$\tau$ reflection symmetry :
 \begin{equation}
U^{\mu\tau}= \left(\begin{array}{ccc}
c_{12}^{\mu\tau} c_{13}^{\mu\tau} &s_{12}^{\mu\tau} c_{13}^{\mu\tau} &s_{13}^{\mu\tau}\\
\frac{1}{\sqrt{2}}( c_{12}^{\mu\tau} s_{13}^{\mu\tau} \pm is_{12}^{\mu\tau})&\frac{1}{\sqrt{2}}(s_{12}^{\mu\tau}s_{13}^{\mu\tau}\mp ic_{12}^{\mu\tau})&\frac{1}{\sqrt{2}}c_{13}^{\mu\tau}\\
\frac{1}{\sqrt{2}}(c_{12}^{\mu\tau}s_{13}^{\mu\tau} \mp i s_{12}^{\mu\tau})&\frac{1}{\sqrt{2}}(s_{12}^{\mu\tau}s_{13}^{\mu\tau}\pm i c_{12}^{\mu\tau})&\frac{1}{\sqrt{2}}c_{13}^{\mu\tau}
\end{array}\right).\label{eq:16}
\end{equation}
We see that the first row of $U^{\mu\tau}$ is entirely real, while the second and third rows are related through complex conjugation. Therefore, the mixing matrix satisfies the Harrison-Scott form given in Eq.~(\ref{eq:13}). The upper $'+'$ and lower  $'-'$ signs in Eq.~(\ref{eq:16}) correspond to Case-I and Case-II respectively. With the reflection symmetric mixing matrix in Eq.~(\ref{eq:16}), the mass matrix $M_\nu^{\mu\tau}$ can be diagonalized as
\begin{equation}
\left( U^{\mu\tau}\right)^\dagger M_\nu^{\mu\tau}\left( U^{\mu\tau}\right)^*=D_\nu^{\mu\tau} = Diag({m_1^{\mu\tau}, m_2^{\mu\tau}, m_3^{\mu\tau})},\label{eq:17}
\end{equation}
where $m_1^{\mu\tau}$, $m_2^{\mu\tau}$ and $m_3^{\mu\tau}$ denote the neutrino mass eigenvalues in the $\mu$--$\tau$ reflection symmetric limit.
\indent~ We now present the expressions for the lepton mixing angles and CP phases in terms of the elements of $U$. The mixing angles are given by
\begin{equation}
\sin^2\theta_{13}= \vert U_{e3} \vert^2, \ \ 
          \sin^2\theta_{23}= \frac{\vert U_{\mu3} \vert^2}{1-\vert U_{e3} \vert^2}, \ \
                 \sin^2\theta_{12}= \frac{\vert U_{e2} \vert^2}{1-\vert U_{e3} \vert^2},\label{eq:18}
\end{equation}
while the CP phases are obtained from
\begin{equation}
\displaystyle \delta - \phi_1 = - \arctan \left(\frac{Im (U_{e3})}{Re (U_{e3})}\right),\label{eq:19}
\end{equation}
\begin{equation}
\displaystyle \phi_2 = \arctan \left(\frac{Im (U_{\mu 3})}{Re (U_{\mu 3})}\right), \ \
                       \phi_3 = \arctan \left(\frac{Im (U_{\tau 3})}{Re (U_{\tau 3})}\right).\label{eq:20}
\end{equation}
We also estimate CP violation using the Jarlskog invariant,
\begin{equation}
 J=Im[U_{e2} U_{\mu3}U^*_{e3} U^*_{\mu2}]
,\label{eq:21} 
\end{equation}
where $U_{e2}$,\ $U_{e3}$,\ $U_{\mu2}$ and $U_{\mu3}$ are the elements of the lepton mixing matrix given in Eq.~(\ref{eq:9}). Substituting these elements in Eq.~(\ref{eq:21}), we obtain
\begin{equation}
J = s_{12}c_{12}s_{13} c_{13}^2s_{23}c_{23}  \sin\delta. \label{eq:22}
\end{equation}
It is important to note that neutrinoless double-beta ($0\nu\beta\beta$) decay provides a sensitive probe of the Majorana nature of neutrinos. The rate of this process depends on the effective Majorana neutrino mass, which is given by
\begin{equation}
  |\langle m \rangle_{ee}|=\vert \sum_{i=1}^3 m_i U_{ei}^2 \vert,\label{eq:23} 
\end{equation}
where $m_i$ are the neutrino mass eigenvalues and $U_{ei}$ are the first row elements of the lepton mixing matrix. With the parameterization of the lepton mixing matrix defined in Eq.(\ref{eq:9}), the above expression becomes
\begin{equation}
    |\langle m \rangle_{ee}|=\vert m_1 c_{12}^2 c_{13}^2 e^{2i\rho} + m_2 s_{12}^2 c_{13}^2 e^{2i\sigma} + m_3 s_{13}^2 e^{-2i\delta}\vert.\label{eq:24}
\end{equation}

\section{The minimal seesaw and $\mu$-$\tau$ reflection symmetry}
\indent~ In the Type-I seesaw framework, the effective Majorana neutrino mass matrix in Eq.~(\ref{eq:3}) is given by
\begin{equation}
M_\nu = -M_D M_R^{-1} M_D^T,
\label{eq:25}
\end{equation}
where $M_D$ and $M_R$ are the $3\times 3$ Dirac and heavy right-handed Majorana neutrino mass matrices respectively. The most popular economical version of the Type-I seesaw is the minimal seesaw. In this framework, $M_D$ and $M_R$ in Eq.(\ref{eq:25}) become $3\times2$ and $2\times2$ matrices respectively. To realize $\mu$-$\tau$ reflection symmetry in the minimal seesaw, we adopt the texture of the Dirac mass matrix from Ref.\cite{nnath} which is given by 
\begin{equation}
M_D = \left(
\begin{array}{cc}
a& a^{*} \\
b & c\\
c^{*} & b^{*}
\end{array}
\right), \label{eq:26}
\end{equation}
where $a$,\ $b$ and $c$ are complex elements. We choose the basis where $M_R$ is diagonal, given by
\begin{equation}
M_R = \left(
\begin{array}{cc}
M_1 & 0 \\
0 & M_2\\
\end{array}
\right), \label{eq:27}
\end{equation} 
where $M_1$ and $M_2$ are the heavy right-handed Majorana neutrino masses. Using Eqs.~(\ref{eq:26}) and (\ref{eq:27}) in Eq.~(\ref{eq:25}), the effective light neutrino mass matrix takes the form 
\begin{equation}
M_\nu = 
\left(
\begin{array}{ccc}
\dfrac{a^2}{M_1} + \dfrac{(a^{*})^2}{M_2}
&
\dfrac{ab}{M_1} + \dfrac{a^{*}c}{M_2}
&
\dfrac{ac^{*}}{M_1} + \dfrac{a^{*}b^{*}}{M_2}
\\[2mm]
\dfrac{ab}{M_1} + \dfrac{a^{*}c}{M_2}
&
\dfrac{b^2}{M_1} + \dfrac{c^2}{M_2}
&
\dfrac{bc^{*}}{M_1} + \dfrac{cb^{*}}{M_2}
\\[2mm]
\dfrac{ac^{*}}{M_1} + \dfrac{a^{*}b^{*}}{M_2}
&
\dfrac{bc^{*}}{M_1} + \dfrac{cb^{*}}{M_2}
&
\dfrac{(c^{*})^2}{M_1} + \dfrac{(b^{*})^2}{M_2}
\end{array}
\right).\label{eq:28}
\end{equation}
However, the above mass matrix does not exhibit reflection symmetry. Therefore, in order to preserve reflection symmetry, we consider the approximation $M_1\simeq M_2=M$. With this approximation, the above mass matrix becomes $\mu$-$\tau$ reflection symmetric form given in Eq.~(\ref{eq:5}), with
\begin{equation}
m_{ee} = \frac{a^2 + a^{*2}}{M},\
 m_{e\mu} = \frac{ab + a^*c}{M},\
m_{\mu\mu} = \frac{b^2 + c^2}{M}, \
m_{\mu\tau} = \frac{bc^* + cb^*}{M}.\label{eq:29}
\end{equation}
An important feature of the minimal seesaw framework is that one of the three light neutrinos is massless. Since there exist two scenarios regarding the order of mass eigenvalues- NO, $(m_1<m_2<m_3)$ and IO, $(m_3<m_2<m_1)$, we consider the following two conventional situations:
\begin{equation}
    m_1=0 \label{eq:30}
\end{equation}
for the NO scenario and 
\begin{equation}
    m_3=0, \label{eq:31}
\end{equation}
for the IO scenario. As a consequence of the vanishing nature of one mass eigenvalue in the minimal seesaw framework, one of the Majorana CP phases becomes physically irrelevant. It can be seen from the expression of the effective Majorana neutrino mass in Eq.(\ref{eq:24}). For NO, $m_1=0$ eliminates the Majorana phase $\rho$, leaving $\sigma$ as the only physically relevant Majorana CP phase. Similarly, in IO, $\rho$ can also be eliminated by a global rephasing of the left-handed neutrino fields: $\nu_{\alpha L}\rightarrow e^{-i\rho}\nu_{\alpha L}\ (\alpha=e,\ \mu,\ \tau)$. Under this rephasing, the physically relevant Majorana phase becomes $\sigma^\prime = \sigma - \rho$.
\section{Radiative corrections to $\mu$-$\tau$ reflection symmetry in minimal seesaw}
\indent~ The effective Majorana neutrino mass matrix in Eq.(\ref{eq:25}) can also be expressed as 
\begin{equation}
M_\nu =\kappa v^2\label{eq:32}
\end{equation}
in the SM and 
\begin{equation}
M_\nu =\kappa v^2 \frac{\tan^2\beta}{\left(1+\tan^2\beta \right)}\label{eq:33}
\end{equation}
in the MSSM, where $\kappa$  denotes the coefficient of the effective dimension-five Weinberg operator, given by
\begin{equation}
 O_5^{SM}=-(1/2)\kappa (\bar{l_L}H)(H^T l_L^c)\label{eq:34}
\end{equation}
 and
\begin{equation}
 O_5^{MSSM}=-(1/2)\kappa (\mathbf{LH})(\mathbf{H L}) +\mathrm{h.c}\label{eq:35}
\end{equation}
in the SM and MSSM respectively. In Eqs.(\ref{eq:32}) and (\ref{eq:33}), $v$ denotes the vev of the SM Higgs field, with $<H>=v\simeq 174\ GeV$, while $\tan\beta$ denotes the ratio of the vevs of two Higgs doublets in the MSSM. Further, in Eq.(\ref{eq:34}), $l_L$ represents the SM lepton doublet, whereas in  Eq.(\ref{eq:35}), $\mathbf{L}$ and $\mathbf{H}$ stand for the lepton doublet and up-type Higgs superfield in the MSSM respectively. The one-loop renormalization group equation (RGE) of the Majorana neutrino mass matrix is given by \cite{ chank,babu,casas,antusch,antusch1,chank1,ray}
\begin{equation}
16 \pi^2 \frac{dM_\nu}{dt}= \alpha M_\nu + C \left[ \lbrace(Y_l Y_l^{\dagger})M_\nu
+M_\nu (Y_l Y_l^{\dagger})^T \rbrace \right],\label{eq:36}
\end{equation}
where $t$ stands for $\ln(\mu/\mu_{\circ})$ with $\mu$ being the renormalization scale and $Y_l$ is the charged lepton Yukawa coupling matrix. The constants $C$ and $\alpha$ are given by
\begin{equation}
C=-\frac{3}{2},\ \ \alpha \simeq -3g_2^2 + 6 y_t^2 + \lambda \label{eq:37}
\end{equation}
for SM and 
\begin{equation}
 C= 1, \ \ \ \alpha \simeq -\frac{6}{5}g_1^2 -6 g_2^2 + 6 y_t^2 \label{eq:38}
\end{equation}
for MSSM, with $g_{1,2}$, $y_t$ and $\lambda$ being the gauge coupling constants, top quark Yukawa coupling constant and SM quartic Higgs coupling constant respectively. The integral solution of Eq.~(\ref{eq:36}) is given by \cite{ingr1, ingr2}
\begin{equation}
 M_{\nu}(\Lambda_{EW}) = I_{\alpha} I_l M_{\nu}(\Lambda_{\mu\tau}) I_l. \label{eq:39}
\end{equation}
In the above equation, $M_{\nu}(\Lambda_{EW})$ on the left-hand side denotes the low-energy neutrino mass matrix at $\Lambda_{EW}$, while  $M_{\nu}(\Lambda_{\mu\tau})$ on the right-hand side denotes the $\mu$-$\tau$ reflection symmetric mass matrix defined in Eq.(\ref{eq:5})) at the high-energy scale. Further, the factors appearing on the right-hand side of Eq.~(\ref{eq:39}) are given by
\begin{equation}
  \displaystyle I_\alpha = \exp \left[- \frac{C}{16 \pi^2} 
  \int_{\ln\left(\Lambda_{EW}/1GeV\right)}^{\ln\left(\Lambda_{\mu\tau}/1GeV\right)}\alpha(t)dt \right],\label{eq:40}
 \end{equation}
and $I_l=Diag\{I_e,I_{\mu},I_{\tau}\}$ with
\begin{equation}
   \displaystyle I_{e,\mu,\tau}  = \exp \left[- \frac{C}{16 \pi^2} 
              \int_{\ln\left(\Lambda_{EW}/1GeV\right)}^{\ln\left(\Lambda_{\mu\tau}/1GeV\right)} y_{e,\mu,\tau}^2(t)dt \right].\label{eq:41}
    \end{equation}
Since $y_e^2<<y_{\mu}^2<<y_{\tau}^2$, we approximate $I_e \approx 1$,
$I_{\mu}\approx 1$ and $I_{\tau}\approx 1+\epsilon$, with 
\begin{equation}
 \displaystyle \epsilon  = - \frac{C}{16 \pi^2} 
\int_{\ln\left(\Lambda_{EW}/1GeV\right)}^{\ln\left(\Lambda_{\mu\tau}/1GeV\right)} y_{\tau}^2(t)dt. \label{eq:42}
    \end{equation} 
With these approximations, the low-energy mass matrix in Eq.(\ref{eq:39}) becomes
\begin{equation}
M_{\nu} =I_{\alpha} \left(\begin{array}{ccc}
 m_{ee} & m_{e\mu} & (1+\epsilon) m_{e \mu}^{*}\\
 m_{e\mu} & m_{\mu\mu} & (1+\epsilon)m_{\mu\tau}\\
  (1+\epsilon) m_{e \mu}^{*} & (1+\epsilon)m_{\mu\tau} & (1+\epsilon)^2  m_{\mu\mu}^{*}\\
   \end{array}\right).\label{eq:43}
\end{equation}
In the leading O($\epsilon$), the above mass matrix can be expressed as 
\begin{equation}
    M_\nu \approx I_\alpha \left[ M_\nu^{\mu\tau}
    +\epsilon \Delta M_\nu\right],\label{eq:44}
\end{equation}
where  $M_\nu^{\mu\tau}$ is given in Eq.(\ref{eq:5}) and $\Delta M_\nu $ is given by
\begin{equation}
{\Delta M}_\nu = \left(\begin{array}{ccc}
 0 & 0 &  m_{e\mu}^{*}\\
 0 & 0 & m_{\mu\tau}\\
 ( m_{e\mu}^{*} & m_{\mu\tau} & 2 m_{\mu\mu}^{*} \\
 \end{array}\right).\label{eq:45}
\end{equation}
 The low-energy neutrino mass matrix in Eq.~(\ref{eq:44}) can be diagonalized by the lepton mixing matrix $U$ in Eq.~(\ref{eq:9}), employing the diagonalization relation given in Eq.~(\ref{eq:8}). We can decompose the lepton mixing matrix $U$ as
\begin{equation}
    U= U^{\mu\tau} + \Delta U,\label{eq:46}
\end{equation}
where $U^{\mu\tau}$ denotes the reflection symmetric mixing matrix, which diagonalizes the symmetric mass matrix $M_\nu^{\mu\tau}$ through the relation given in Eq.~(\ref{eq:17}). In  Eqs.(\ref{eq:46}), $\Delta U$ represents the deviation from $U^{\mu\tau}$. Now, using Eqs.~(\ref{eq:8}), (\ref{eq:17}) and (\ref{eq:46}) in Eq.~(\ref{eq:44}), we obtain
\begin{equation}
D_{\nu}\approx I_\alpha\left[ D_{\nu}^{\mu\tau} +\epsilon \left(U^{\mu\tau}\right)^{\dagger}{\Delta M}_\nu \left(U^{\mu\tau}\right)^{*} + D_{\nu}^{\mu\tau} \left(U^{\mu\tau}\right)^{T} (\Delta U)^{*}
 + {\Delta U}^{\dagger}U^{\mu\tau}D_{\nu}^{\mu\tau} \right],\label{eq:47}
\end{equation}
where second order terms in  ${\Delta M}_\nu$ and $\Delta U$ are neglected. The above equation allows us to obtain expressions for the low-energy mass eigenvalues, mixing angles and CP phases. In the conventional Type I seesaw framework, this task has been accomplished in \cite{pegu}. In the present work, we perform the same approach in the minimal seesaw framework, where one of the light neutrino masses is vanishing. Since there exist two scenarios regarding the order of mass eigenvalues, the choice of the vanishing mass eigenvalue and the physical Majorana phase differs in between NO and IO scenarios. As a result, Eq.(\ref{eq:47}) leads to different analytical expressions of the low-energy parameters for the NO and IO scenarios. Accordingly, we derive the analytical expressions for all low-energy neutrino parameters separately for the two scenarios. The relevant treatments are presented in the following subsections.
\subsection{Normal Order: $m_1 < m_2 < m_3$}
\indent~ In the NO scenario $m_1^{\mu\tau}=0$, and consequently, from 
Eq.~(\ref{eq:47}), by the comparison of diagonal elements, we have
\begin{equation}
m_1 \simeq 0,\label{eq:48}
\end{equation}
\begin{equation}
 m_2 \simeq I_{\alpha} \left[ 1 + \epsilon \left[ \left( s_{12}^{\mu\tau}\right)^2 \left( s_{13}^{\mu\tau}\right)^2 + \left( c_{12}^{\mu\tau}\right)^2 \right] \right] m_{2}^{\mu\tau},\label{eq:49}
\end{equation}
\begin{equation}
m_3 \simeq I_{\alpha} \left[ 1 + \epsilon \left( c_{13}^{\mu\tau}\right)^2  \right] m_{3}^{\mu\tau}.\label{eq:50}
\end{equation}
The left-hand sides of Eqs.~(\ref{eq:48})--(\ref{eq:50}) represent the low-energy mass eigenvalues, and the right-hand sides express them in terms of the high-energy parameters. Equation~(\ref{eq:48}) shows that 
$m_1$ remains massless at the electroweak scale under one-loop radiative corrections. On the other hand, the remaining two mass eigenvalues, $m_2$ and $m_3$, receive non-zero radiative contributions through the factors $I_\alpha$ and $\epsilon$, as shown in Eqs.~(\ref{eq:49}) and (\ref{eq:50}).\\
\indent~ Next, we obtain the analytical expressions for the low-energy mixing angles and CP phases. To do this, we first determine the low-energy mixing matrix $U$ from Eq.~(\ref{eq:46}). This equation implies $U$ is composed of two matrices, $U^{\mu\tau}$ and $\Delta U$. The matrix $U^{\mu\tau}$ is given in Eq.~(\ref{eq:16}), while $\Delta U$ can be determined from Eq.~(\ref{eq:47}). By comparing the off-diagonal elements of Eq.~(\ref{eq:47}), we obtain the elements of $\Delta U$ as follows:
\begin{equation}
\displaystyle \Delta U_{e1}=\frac{\epsilon}{2}\left[  \left(s_{12}^{\mu\tau}\right)^2 \left(c_{13}^{\mu\tau}\right)^2  + \left(s_{13}^{\mu\tau}\right)^2 \right] c_{12}^{\mu\tau}  c_{13}^{\mu\tau},\label{eq:51}
\end{equation}
\begin{equation}
\displaystyle \Delta U_{e2}=\frac{\epsilon}{2}\left[ \left[ \left(s_{13}^{\mu\tau}\right)^2 \Delta_{32} - \left(c_{12}^{\mu\tau}\right)^2 \left(c_{13}^{\mu\tau}\right)^2 \right] 
  s_{12}^{\mu\tau} \mp i \left[ \Delta_{32}^{-1}+1
\right]  c_{12}^{\mu\tau} s_{13}^{\mu\tau}\right]c_{13}^{\mu\tau},\label{eq:52}
\end{equation}
\begin{equation}
\displaystyle \Delta U_{e3}= -\frac{\epsilon}{2}\left[ \left[ \left(c_{12}^{\mu\tau}\right)^2  + \left(s_{12}^{\mu\tau}\right)^2 \Delta_{32}\right] s_{13}^{\mu\tau}
 \mp i \left[ 1
- \Delta_{32}^{-1}\right] s_{12}^{\mu\tau} c_{12}^{\mu\tau}\right] \left(c_{13}^{\mu\tau}\right)^2,\label{eq:53}
\end{equation}
\begin{equation}
\begin{aligned}
\displaystyle \Delta U_{\mu 3} = & \frac{\epsilon}{2\sqrt{2}}
 \bigl[  \left(s_{13}^{\mu\tau}\right)^2 \left[ \left(c_{12}^{\mu\tau}\right)^2  
  + \left(s_{12}^{\mu\tau}\right)^2 \Delta_{32} \right]
 -\left[ \left(s_{12}^{\mu\tau}\right)^2    
  + \left(c_{12}^{\mu\tau}\right)^2 \Delta_{32}^{-1} \right] \\
  & \mp i \left[ 2  - \Delta_{32}
   -\Delta_{32}^{-1}\right] s_{12}^{\mu\tau}c_{12}^{\mu\tau}s_{13}^{\mu\tau} \bigr] c_{13}^{\mu\tau} ,
\end{aligned}\label{eq:54}
\end{equation}
\begin{equation}
\begin{aligned}
\displaystyle \Delta U_{\tau 3}= & \frac{\epsilon}{2\sqrt{2}} \bigl[  \left(s_{13}^{\mu\tau}\right)^2 \left[ \left(c_{12}^{\mu\tau}\right)^2 
  + \left(s_{12}^{\mu\tau}\right)^2 \Delta_{32} \right] + \left[ \left(s_{12}^{\mu\tau}\right)^2  + \left(c_{12}^{\mu\tau}\right)^2 \Delta_{32}^{-1} \right] \\
 & \mp i \left[  \Delta_{32} -\Delta_{32}^{-1}\right] s_{12}^{\mu\tau}c_{12}^{\mu\tau}s_{13}^{\mu\tau} \bigr] c_{13}^{\mu\tau},
\end{aligned}\label{eq:55}
\end{equation}
in the NO scenario. In the above equations, 
\begin{equation}
 \displaystyle\Delta_{32}= \frac{m_3^{\mu\tau} + m_2^{\mu\tau}}{m_3^{\mu\tau} - m_2^{\mu\tau}}, \label{eq:56}
\end{equation}
and the $'+'$ and $'-'$ signs on right-hand side correspond to Case-I $(\delta^{\mu\tau}= \frac{\pi}{2})$ and Case-II $(\delta^{\mu\tau}= \frac{3\pi}{2})$ respectively.  Note that, in the above, we present only the relevant elements of $\Delta U$ required for the present analysis. These elements are sufficient to derive the analytical expressions for the low-energy mixing angles and CP phases. Substituting the above expressions for $\Delta U$, together with $U^{\mu\tau}$, into Eq.~(\ref{eq:46}), we obtain the elements of the low-energy mixing matrix $U$ as
\begin{equation}
\displaystyle U_{e1}=c_{12}^{\mu\tau} c_{13}^{\mu\tau}+\frac{\epsilon}{2}\left[ \left(s_{12}^{\mu\tau}\right)^2 \left(c_{13}^{\mu\tau}\right)^2  + \left(s_{13}^{\mu\tau}\right)^2 \right] c_{12}^{\mu\tau}  c_{13}^{\mu\tau},\label{eq:57}
\end{equation}
\begin{equation}
\displaystyle U_{e2}=s_{12}^{\mu\tau} c_{13}^{\mu\tau}+\frac{\epsilon}{2}\left[ \left[ \left(s_{13}^{\mu\tau}\right)^2 \Delta_{32} - \left(c_{12}^{\mu\tau}\right)^2 \left(c_{13}^{\mu\tau}\right)^2 \right] 
  s_{12}^{\mu\tau} \mp i \left[ \Delta_{32}^{-1}+1
\right]  c_{12}^{\mu\tau} s_{13}^{\mu\tau}\right]c_{13}^{\mu\tau},\label{eq:58}
\end{equation}
\begin{equation}
\displaystyle U_{e3}=s_{13}^{\mu\tau} -\frac{\epsilon}{2}\left[ \left[ \left(c_{12}^{\mu\tau}\right)^2 + \left(s_{12}^{\mu\tau}\right)^2 \Delta_{32}\right] s_{13}^{\mu\tau}
 \mp i \left[ 1
- \Delta_{32}^{-1}\right] s_{12}^{\mu\tau} c_{12}^{\mu\tau}\right] \left(c_{13}^{\mu\tau}\right)^2,\label{eq:59}
\end{equation}
\begin{equation}
\begin{aligned}
\displaystyle U_{\mu 3} = & \frac{c_{13}^{\mu\tau}}{\sqrt{2}} + \frac{\epsilon}{2\sqrt{2}}
 \bigl[  \left(s_{13}^{\mu\tau}\right)^2 \left[ \left(c_{12}^{\mu\tau}\right)^2  
  + \left(s_{12}^{\mu\tau}\right)^2 \Delta_{32} \right]
 -\left[ \left(s_{12}^{\mu\tau}\right)^2   
  + \left(c_{12}^{\mu\tau}\right)^2 \Delta_{32}^{-1} \right] \\
  & \mp i \left[ 2  - \Delta_{32}
  -\Delta_{32}^{-1}\right] s_{12}^{\mu\tau}c_{12}^{\mu\tau}s_{13}^{\mu\tau} \bigr] c_{13}^{\mu\tau},
\end{aligned}\label{eq:60}
\end{equation}
\begin{equation}
\begin{aligned}
\displaystyle U_{\tau 3}= &  \frac{c_{13}^{\mu\tau}}{\sqrt{2}}+\frac{\epsilon}{2\sqrt{2}} \bigl[  \left(s_{13}^{\mu\tau}\right)^2 \left[ \left(c_{12}^{\mu\tau}\right)^2 
  + \left(s_{12}^{\mu\tau}\right)^2 \Delta_{32} \right] + \left[ \left(s_{12}^{\mu\tau}\right)^2 \Delta_{31}^{-1} + \left(c_{12}^{\mu\tau}\right)^2 \Delta_{32}^{-1} \right] \\
 & \mp i \left[ \Delta_{32} -\Delta_{32}^{-1}\right] s_{12}^{\mu\tau}c_{12}^{\mu\tau}s_{13}^{\mu\tau} \bigr] c_{13}^{\mu\tau}.
\end{aligned}\label{eq:61}
\end{equation}
Finally, substituting the above expressions into Eq.~(\ref{eq:18}), we obtain the corresponding expressions for the low-energy mixing angles in the NO scenario within the minimal seesaw framework: 
\begin{equation}
\displaystyle \sin^2\theta_{13} \approx \left(s_{13}^{\mu\tau}\right)^2 - \epsilon \left(s_{13}^{\mu\tau}\right)^2 c_{13}^{\mu\tau} \left[\left(c_{12}^{\mu\tau}\right)^2  + \left(s_{12}^{\mu\tau}\right)^2 \Delta_{32}\right],\label{eq:62}
 \end{equation}
\begin{equation}
\displaystyle \sin^2\theta_{23} \approx \frac{1}{2} -\frac{\epsilon}{2}\left[ \left(s_{12}^{\mu\tau}\right)^2 +\left(c_{12}^{\mu\tau}\right)^2 \Delta_{32}^{-1}\right],\label{eq:63}
\end{equation}
\begin{equation}
\displaystyle \sin^2\theta_{12} \approx \left(s_{12}^{\mu\tau}\right)^2 - \epsilon \left(s_{12}^{\mu\tau}\right)^2 c_{12}^{\mu\tau} \left[ \left(c_{13}^{\mu\tau}\right)^2 -\left(s_{13}^{\mu\tau}\right)^2 \Delta_{32}\right].
\label{eq:64}
\end{equation}
In the above expressions, we neglect the higher-order terms dominated by $O(\epsilon^2)$. All the parameters on the right-hand sides of the above equations correspond to the flavor symmetry scale, where the reflection symmetry is assumed to be preserved. \\
\indent~ Turning to the prediction of CP violation in the present treatment, we first obtain the analytical expressions for the Jarlskog invariant $J$ and the effective Majorana neutrino mass $|\langle m \rangle_{ee}|$ from Eqs.~(\ref{eq:21}) and (\ref{eq:23}) respectively. The resulting expressions for $J$ and $|\langle m \rangle_{ee}|$ are then used to derive the analytical expressions for the Dirac CP phase and the Majorana CP phases respectively. The  expression for the low-energy Jarlskog invariant in the leading order of $\epsilon$ is given by 
\begin{equation}
 \displaystyle J\simeq \pm J_0^{\mu\tau}\ \mp \epsilon J_1^{\mu\tau},
  \label{eq:65}
\end{equation}
where 
\begin{equation}
   J_0^{\mu\tau}= \frac{1}{2}s_{12}^{\mu\tau}s_{13}^{\mu\tau}\left(c_{13}^{\mu\tau}\right)^2 c_{12}^{\mu\tau},\label{eq:66}
\end{equation}
and 
\begin{equation}
\begin{aligned}
    J_1^{\mu\tau}&= \frac{1} {2} \bigl[ \frac{1}{\sqrt{2}} \left(\left(s_{13}^{\mu\tau}\right)^2  -\left[ 1 + \left(s_{12}^{\mu\tau}\right)^2\right]\left(c_{13}^{\mu\tau}\right)^2 \right)\\& - \left[ 1
- \Delta_{32}^{-1}\right]\left(s_{12}^{\mu\tau}\right)^2 \left(c_{13}^{\mu\tau}\right)^2 - \frac{1}{\sqrt{2}}\left[ 2  - \Delta_{32} -\Delta_{32}^{-1}\right]c_{12}^{\mu\tau} s_{13}^{\mu\tau} s_{12}^{\mu\tau} \\& - \frac{1}{2} \left[ \Delta_{32}^{-1}\right]
s_{13}^{\mu\tau} \bigr] s_{12}^{\mu\tau}\left(c_{13}^{\mu\tau}\right)^2 s_{13}^{\mu\tau} c_{12}^{\mu\tau}.
    \end{aligned}\label{eq:67}
\end{equation}
 Similarly, the low-energy expression for the effective Majorana neutrino mass is given by 
\begin{equation}
   |\langle m \rangle_{ee}|\simeq I_\alpha \vert \left(\langle m \rangle_{ee} \right)_0^{\mu\tau}  + \epsilon \ \left(\langle m \rangle_{ee} \right)_1^{\mu\tau} \vert,\label{eq:68}
\end{equation}
where
\begin{equation}
   \left(\langle m \rangle_{ee} \right)_0^{\mu\tau}=  \left(s_{12}^{\mu\tau}\right)^2 \left(c_{13}^{\mu\tau}\right)^2 m_2^{\mu\tau} + \left(s_{13}^{\mu\tau}\right)^2m_3^{\mu\tau} \label{eq:69}
\end{equation}
and 
\begin{equation}
    \begin{aligned}
        \left(\langle m \rangle_{ee} \right)_1^{\mu\tau}= & \left(s_{13}^{\mu\tau}\right)^2 \left(c_{13}^{\mu\tau}\right)^2\left[ \left(s_{12}^{\mu\tau}\right)^4 m_2^{\mu\tau} + m_3^{\mu\tau}\right]\\&
       + \left(c_{12}^{\mu\tau}\right)^2 \left(c_{13}^{\mu\tau}\right)^2\left[ \left(s_{12}^{\mu\tau}\right)^2 m_2^{\mu\tau}-\left(s_{13}^{\mu\tau}\right)^2m_3^{\mu\tau} \right]\\&
       + \left(s_{12}^{\mu\tau}\right)^2 \left(c_{13}^{\mu\tau}\right)^2\left[ \left(s_{13}^{\mu\tau}\right)^2\left( m_2^{\mu\tau}- \left(s_{12}^{\mu\tau}\right)^2 m_3^{\mu\tau}     \right)\Delta_{32}\right]\\&
       + \left(c_{12}^{\mu\tau}\right)^2 \left(c_{13}^{\mu\tau}\right)^2\left[\left(s_{12}^{\mu\tau}\right)^2  \left(c_{13}^{\mu\tau}\right)^2m_2^{\mu\tau}\right].
    \end{aligned}\label{eq:70}
\end{equation}
Then using the low-energy expressions of mixing angles and $J$ from Eqs.(\ref{eq:62})--(\ref{eq:64}) and  (\ref{eq:65}) in Eq.~(\ref{eq:22}), a detailed algebraic calculation leads to the expression for the Dirac CP phase:
\begin{equation}
  \displaystyle   \delta^I/\delta^{II}=(\pi/2)/(3\pi/2)\ \mp \epsilon \left(\frac{J_1^{\mu\tau}+ J_2^{\mu\tau}}{J_0^{\mu\tau}} \right),\label{eq:71}
\end{equation}
with
\begin{equation}
\begin{aligned}
\displaystyle J_2^{\mu\tau}=&  \frac{1}{4} \bigl[  s_{12}^{\mu\tau} c_{12}^{\mu\tau} c_{13}^{\mu\tau}\left(\left(c_{12}^{\mu\tau}\right)^2+ \left(s_{12}^{\mu\tau}\right)^2 \Delta_{32}^{-1}\right) \left(\left(c_{13}^{\mu\tau}\right)^2-2 \left(s_{13}^{\mu\tau}\right)^2
\right)\\&
+ s_{12}^{\mu\tau} s_{13}^{\mu\tau} \left(c_{13}^{\mu\tau}\right)^2 \left( \left(c_{13}^{\mu\tau}\right)^2-\left(s_{13}^{\mu\tau}\right)^2\Delta_{32}
\right) \left(\left(c_{12}^{\mu\tau}\right)^2- \left(s_{12}^{\mu\tau}\right)^2
\right)\bigr], \end{aligned}\label{eq:72}
\end{equation}
and $J_0^{\mu\tau}$ and $J_1^{\mu\tau}$ are given in Eqs.(\ref{eq:66}) and (\ref{eq:67}) respectively. Similarly, using the low-energy expression for $|\langle m \rangle_{ee}|$ from Eq.~(\ref{eq:68}), and those of mixing angles and Dirac CP phase $\delta$ in Eq.~(\ref{eq:24}), we obtain the expression for the low-energy Majorana CP phase as
\begin{equation}
\displaystyle\sigma^I/\sigma^{II}=(3\pi/2)/(\pi/2) + \epsilon \left[
\frac{ \left(\langle m \rangle_{ee} \right)_1^{\mu\tau}}{2 \left(\langle m \rangle_{ee} \right)_0^{\mu\tau}} \pm \left(\frac{J_1^{\mu\tau}+ J_2^{\mu\tau}}{J_0^{\mu\tau}} \right)\right]. \label{eq:73}
\end{equation}
 The low-energy expressions for unphysical phases are derived from Eqs.(\ref{eq:19}) and (\ref{eq:20}): 
\begin{equation}
\begin{aligned}
    \displaystyle \phi_1^I/\phi_1^{II}&= (\pi/2)/(3\pi/2)\ \mp \epsilon \left(\frac{J_1^{\mu\tau}+ J_2^{\mu\tau}}{J_0^{\mu\tau}} \right) + \\& \arctan \left[\frac{\mp \frac{\displaystyle\epsilon}{2}\left[\Delta_{32}^{-1}-
 1\right]s_{12}^{\mu\tau} c_{12}^{\mu\tau}\left(c_{13}^{\mu\tau}\right)^2 }{s_{13}^{\mu\tau}- \frac{\displaystyle\epsilon}{2} \left[ \left(s_{12}^{\mu\tau}\right)^2 \Delta_{32}+
\left(c_{12}^{\mu\tau}\right)^2 \right] s_{13}^{\mu\tau}\left(c_{13}^{\mu\tau}\right)^2}  \right], \label{eq:74}
\end{aligned}
\end{equation}
\begin{equation}
\displaystyle \phi_2^I/\phi_2^{II} = \arctan \left[\frac{\pm \frac{\displaystyle\epsilon}{2}\left[\Delta_{32}-
 2+\Delta_{32}^{-1}-\right]s_{12}^{\mu\tau} c_{12}^{\mu\tau}s_{13}^{\mu\tau}}{1+ \frac{\displaystyle\epsilon}{2} \left[ \left(s_{12}^{\mu\tau}\right)^2 \Delta_{32}+
\left(c_{12}^{\mu\tau}\right)^2  \right] \left(s_{13}^{\mu\tau}\right)^2-\left[ \left(c_{12}^{\mu\tau}\right)^2 \Delta_{32}^{-1}+
\left(s_{12}^{\mu\tau}\right)^2  \right]}  \right],\label{eq:75}
\end{equation}
\begin{equation}
\displaystyle \phi_3^I/\phi_3^{II}=  \arctan \left[\frac{\mp \frac{\displaystyle\epsilon}{2}\left[\Delta_{32}-\Delta_{32}^{-1}\right]s_{12}^{\mu\tau} c_{12}^{\mu\tau}s_{13}^{\mu\tau}}{1+ \frac{\displaystyle\epsilon}{2} \left[ \left(s_{12}^{\mu\tau}\right)^2 \Delta_{32}+
\left(c_{12}^{\mu\tau}\right)^2  \right] \left(s_{13}^{\mu\tau}\right)^2+\left[ \left(c_{12}^{\mu\tau}\right)^2 \Delta_{32}^{-1}+
\left(s_{12}^{\mu\tau}\right)^2 \right]}  \right],\label{eq:76}
\end{equation} 
where relevant expressions from Eqs.(\ref{eq:59})--(\ref{eq:61}) have been used. In the above expressions $'+'$ and $'-'$ correspond to Case-I and Case-II respectively.\\
\subsection{Inverted Order: $m_3 < m_1 < m_2$}
\indent~ Using a similar algebraic approach as carried in the NO scenario, we derive the low-energy expressions for all the parameters in IO scenario by setting $m_3^{\mu\tau}=0$. From Eq.~(\ref{eq:47}), we obtain the low-energy mass eigenvalues as
\begin{equation}
m_3 \simeq 0,\label{eq:77}
\end{equation}
\begin{equation}
m_1 \simeq I_{\alpha} \left[ 1 + \epsilon \left[ \left( c_{12}^{\mu\tau}\right)^2 \left( s_{13}^{\mu\tau}\right)^2 +\left( s_{12}^{\mu\tau}\right)^2 \right] \right] m_{1}^{\mu\tau},\label{eq:78}
\end{equation}
\begin{equation}
 m_2 \simeq I_{\alpha} \left[ 1 + \epsilon \left[ \left( s_{12}^{\mu\tau}\right)^2 \left( s_{13}^{\mu\tau}\right)^2 + \left( c_{12}^{\mu\tau}\right)^2 \right] \right] m_{2}^{\mu\tau}.\label{eq:79}
\end{equation}
Right-hand sides of above equation express the low-energy masses in terms of the high-energy parameters. The lowest mass $m_3$ receives no contribution from radiative corrections and remains approximately massless at the electroweak scale.\\
\indent~ Next, comparing the off-diagonal elements of Eq.~(\ref{eq:47}), we obtain the elements of $\Delta U$ as
\begin{equation}
\displaystyle \Delta U_{e1}=\frac{\epsilon}{2}\left[ \left[ \left(s_{12}^{\mu\tau}\right)^2 \left(c_{13}^{\mu\tau}\right)^2 \Delta_{21} -\left(s_{13}^{\mu\tau}\right)^2 \right] c_{12}^{\mu\tau} \mp i \left[ 1+\Delta_{21}^{-1}\right] s_{12}^{\mu\tau} s_{13}^{\mu\tau}\right] c_{13}^{\mu\tau},\label{eq:80}
\end{equation}
\begin{equation}
\displaystyle \Delta U_{e2}=\frac{\epsilon}{2}\left[ \left[ -\left(s_{13}^{\mu\tau}\right)^2  - \left(c_{12}^{\mu\tau}\right)^2 \left(c_{13}^{\mu\tau}\right)^2 \Delta_{21}\right] 
  s_{12}^{\mu\tau} \mp i \left[ \Delta_{21}^{-1}-1
\right]  c_{12}^{\mu\tau} s_{13}^{\mu\tau}\right]c_{13}^{\mu\tau},\label{eq:81}
\end{equation}
\begin{equation}
\displaystyle \Delta U_{e3}= \frac{\epsilon}{2}  s_{13}^{\mu\tau} \left(c_{13}^{\mu\tau}\right)^2,\label{eq:82}
\end{equation}
\begin{equation}
\displaystyle \Delta U_{\mu 3} =  \frac{\epsilon}{2\sqrt{2}} \left(c_{13}^{\mu\tau}\right)^3,
\label{eq:83}
\end{equation}
\begin{equation}
\displaystyle \Delta U_{\tau 3}= \frac{\epsilon}{2\sqrt{2}} 
\left[-\left(s_{13}^{\mu\tau}\right)^2-1 \right] c_{13}^{\mu\tau},
\label{eq:84}
\end{equation}
where
\begin{equation}
    \displaystyle \Delta_{21}= \frac{m_2^{\mu\tau} + m_1^{\mu\tau}}{m_2^{\mu\tau} - m_1^{\mu\tau}}.\label{eq:85}
\end{equation}
In Eq.(\ref{eq:80}) and (\ref{eq:81}), $'+'$ and $'-'$ signs  correspond to Case-I and Case-II respectively. Using the above expressions in Eq.~(\ref{eq:46}), the elements of the low-energy mixing matrix can be obtained as
\begin{equation}
\displaystyle U_{e1}=c_{12}^{\mu\tau} c_{13}^{\mu\tau}+\frac{\epsilon}{2}\left[ \left[ \left(s_{12}^{\mu\tau}\right)^2 \left(c_{13}^{\mu\tau}\right)^2 \Delta_{21} -\left(s_{13}^{\mu\tau}\right)^2 \right] c_{12}^{\mu\tau} \mp i \left[ 1+ \Delta_{21}^{-1}\right] s_{12}^{\mu\tau} s_{13}^{\mu\tau}\right] c_{13}^{\mu\tau},\label{eq:86}
\end{equation}
\begin{equation}
\displaystyle U_{e2}=s_{12}^{\mu\tau} c_{13}^{\mu\tau}+\frac{\epsilon}{2}\left[ \left[ -\left(s_{13}^{\mu\tau}\right)^2 - \left(c_{12}^{\mu\tau}\right)^2 \left(c_{13}^{\mu\tau}\right)^2 \Delta_{21}\right] 
  s_{12}^{\mu\tau} \mp i \left[ -1 +\Delta_{21}^{-1}
\right]  c_{12}^{\mu\tau} s_{13}^{\mu\tau}\right]c_{13}^{\mu\tau},\label{eq:87}
\end{equation}
\begin{equation}
\displaystyle U_{e3}=s_{13}^{\mu\tau} +\frac{\epsilon}{2}  s_{13}^{\mu\tau}\left(c_{13}^{\mu\tau}\right)^2,\label{eq:88}
\end{equation}
\begin{equation}
\begin{aligned}
\displaystyle U_{\mu 3} =  \frac{c_{13}^{\mu\tau}}{\sqrt{2}} + \frac{\epsilon}{2\sqrt{2}}
  \left(c_{13}^{\mu\tau}\right)^3
\end{aligned}\label{eq:89}
\end{equation}
\begin{equation}
\begin{aligned}
\displaystyle U_{\tau 3}= \frac{c_{13}^{\mu\tau}}{\sqrt{2}}+\frac{\epsilon}{2\sqrt{2}}  \left[  -\left(s_{13}^{\mu\tau}\right)^2 -1\right] 
   c_{13}^{\mu\tau}.
\end{aligned}\label{eq:90}
\end{equation}
Finally, employing the above expressions in Eq.~(\ref{eq:18}), we obtain the low-energy mixing angles as
\begin{equation}
\displaystyle \sin^2\theta_{13} \approx \left(s_{13}^{\mu\tau}\right)^2 - \epsilon \left(s_{13}^{\mu\tau}\right)^2 c_{13}^{\mu\tau} ,\label{eq:91}
 \end{equation}
\begin{equation}
\displaystyle \sin^2\theta_{23} \approx \frac{1}{2} + \frac{\epsilon}{2},\label{eq:92}
\end{equation}
\begin{equation}
\displaystyle \sin^2\theta_{12} \approx \left(s_{12}^{\mu\tau}\right)^2 - \epsilon \left(s_{12}^{\mu\tau}\right)^2 c_{12}^{\mu\tau} \left(c_{13}^{\mu\tau}\right)^2 \Delta_{21},
\label{eq:93}
\end{equation}
where we neglect the terms dominated by $O(\epsilon^2)$. All the parameters on the right-hand sides also correspond to the high-energy scale. \\
\indent~ Turning to the measure of CP violation, we derive the low-energy expression for the Jarlskog invariant from Eq.~(\ref{eq:21}). The resulting expression has the same form as Eq.~(\ref{eq:65}), with $J_1^{\mu\tau}$ given by
\begin{equation}
\begin{aligned}
 \displaystyle J_1^{\mu\tau}= & \frac{1}{2\sqrt{2}} \left[\Delta_{21}^{-1}\left(s_{13}^{\mu\tau}\right)^2  -\left[ -1 + \Delta_{21}\left(s_{12}^{\mu\tau}\right)^2\right]\left(c_{13}^{\mu\tau}\right)^2 \right] \\ &  
 - \frac{1}{4} \left[ -1 +\Delta_{21}^{-1} \right]
s_{13}^{\mu\tau}  s_{12}^{\mu\tau}\left(c_{13}^{\mu\tau}\right)^2 s_{13}^{\mu\tau} c_{12}^{\mu\tau}.
\end{aligned}\label{eq:94}
\end{equation}
Similarly, we obtain the low-energy expression for the effective Majorana neutrino mass, which has also the same form as Eq.~(\ref{eq:68}), but with $  \left(\langle m \rangle_{ee} \right)_0^{\mu\tau}$ and $ \left(\langle m \rangle_{ee} \right)_1^{\mu\tau}$ given by  
\begin{equation}
    \left(\langle m \rangle_{ee} \right)_0^{\mu\tau}= \left(c_{12}^{\mu\tau}\right)^2 \left(c_{13}^{\mu\tau}\right)^2 m_1^{\mu\tau}+ \left(s_{12}^{\mu\tau}\right)^2 \left(c_{13}^{\mu\tau}\right)^2 m_2^{\mu\tau},\label{eq:95}
\end{equation}
and 
\begin{equation}
    \begin{aligned}
        \left(\langle m \rangle_{ee} \right)_1^{\mu\tau} = & \left(s_{13}^{\mu\tau}\right)^2 \left(c_{13}^{\mu\tau}\right)^2\left[\left(c_{12}^{\mu\tau}\right)^4 m_1^{\mu\tau} + \left(s_{12}^{\mu\tau}\right)^4 m_2^{\mu\tau} \right]\\&
       + \left(c_{12}^{\mu\tau}\right)^2 \left(c_{13}^{\mu\tau}\right)^2\left[ \left(s_{12}^{\mu\tau}\right)^2\left( m_1^{\mu\tau}+ m_2^{\mu\tau}     \right)\right]\\&
       + \left(s_{12}^{\mu\tau}\right)^2 \left(c_{13}^{\mu\tau}\right)^2\left[\left(c_{12}^{\mu\tau}\right)^2 \left(c_{13}^{\mu\tau}\right)^2 m_1^{\mu\tau} \Delta_{21} - \left(s_{13}^{\mu\tau}\right)^2 m_2^{\mu\tau}\right]\\&
       + \left(c_{12}^{\mu\tau}\right)^2 \left(c_{13}^{\mu\tau}\right)^2\left[-\left(s_{13}^{\mu\tau}\right)^2 m_1^{\mu\tau} -\left(s_{12}^{\mu\tau}\right)^2  \left(c_{13}^{\mu\tau}\right)^2m_2^{\mu\tau} \Delta_{21}\right].\label{eq:96}
    \end{aligned}
\end{equation}
Finally, using the expressions for $J$ and $|\langle m \rangle_{ee}|$ derived above, we obtain the low-energy expressions for the Dirac and Majorana CP phases from Eqs.~(\ref{eq:22}) and (\ref{eq:24}), respectively. The analytical expression for the Dirac CP phase is given by 
\begin{equation}
  \displaystyle   \delta^I/\delta^{II}=(\pi/2)/(3\pi/2)\ \mp \epsilon \left(\frac{J_1^{\mu\tau}+ J_2^{\mu\tau}}{J_0^{\mu\tau}} \right),\label{eq:97}
\end{equation}
where
\begin{equation}
\begin{aligned}
\displaystyle J_2^{\mu\tau}=&  \frac{1}{2} \left[  \left(s_{12}^{\mu\tau}\right)^2 s_{13}^{\mu\tau} \left(c_{13}^{\mu\tau}\right)^4 \Delta_{21}-s_{13}^{\mu\tau} \left(c_{12}^{\mu\tau}\right)^2\left(c_{13}^{\mu\tau}\right)^4\Delta_{21}- 2 \left(s_{13}^{\mu\tau}\right)^3 c_{13}^{\mu\tau}s_{12}^{\mu\tau}c_{12}^{\mu\tau}\right] \end{aligned}\label{eq:98}
\end{equation}
and $J_0^{\mu\tau}$ and $J_1^{\mu\tau}$ are given in Eqs.~(\ref{eq:66}) and (\ref{eq:94}), respectively.
 The expression for the physically relevant Majorana phase in the IO is given by
\begin{equation}
\displaystyle(\sigma')^I/(\sigma')^{II}=(\sigma^I/\sigma^{II})-(\rho^I/\rho^{II})=\pm \epsilon\left[1+\frac{1}{2}\left((s_{13}^{\mu\tau})^2-\frac{ \left(\langle m \rangle_{ee} \right)_1^{\mu\tau}}{ \left(\langle m \rangle_{ee} \right)_0^{\mu\tau}} \right) \right],\label{eq:99}
\end{equation}
where $ \left(\langle m \rangle_{ee} \right)_0^{\mu\tau}$ and $ \left(\langle m \rangle_{ee} \right)_1^{\mu\tau}$ are defined in Eqs.(\ref{eq:95}) and (\ref{eq:96}). The analytical expressions for the unphysical phases are obtained using Eqs.~(\ref{eq:19}) and (\ref{eq:20}). However, unlike the NO scenario, the expression for $U_{e3}$ in Eq.~(\ref{eq:88}) contains no imaginary term. Consequently, the right-hand side of Eq.~(\ref{eq:19}) vanishes, leading to
\begin{equation}
\phi_1^{I}/\phi_1^{II} = \delta^{I}/\delta^{II}.\label{eq:100}
\end{equation} 
Similarly, as the expressions for $U_{\mu3}$ and $U_{\tau3}$ in Eqs.~(\ref{eq:89}) and (\ref{eq:90}) contain no imaginary terms, Eq.~(\ref{eq:20}) yields
\begin{equation}
\phi_2^{I}/\phi_2^{II} = \phi_3^{I}/\phi_3^{II} = 0.
\label{eq:103}
\end{equation}

\section{Numerical analysis and results}
\indent~ In this section, we perform a numerical analysis based on the analytical expressions derived in the previous section. All these expressions enable us to estimate the low-energy values of neutrino parameters in a straightforward way once the high-energy parameters on the right-hand sides of these expressions are taken as inputs.  We choose the low-energy scale at the top quark mass scale: $\Lambda_{EW}=m_t=172.76\ GeV$, while the high-energy flavor symmetry scale is set at $\Lambda_{\mu\tau}=10^{14}\ GeV$. The high-energy input values of the atmospheric mixing angle and CP phases are constrained by the reflection symmetry at maximal values. As discussed in Section~2, this symmetry allows two possible values of the Dirac CP phase. Accordingly, we consider two separate cases: Case-I with $\delta^{\mu\tau}=\pi/2$ and Case-II with $\delta^{\mu\tau}=3\pi/2$, as presented after Eq.~(\ref{eq:13}). Thus the input values of CP phases are taken as per Eqs.~(\ref{eq:14}) and (\ref{eq:15}). Apart from the input values of $\theta_{23}^{\mu\tau}$ and CP phases, the remaining high-energy parameters, viz., mass eigenvalues, $\theta_{12}^{\mu\tau}$ and $\theta_{13}^{\mu\tau}$, act as free parameters in the analysis. The input values of these free parameters are to be chosen in such a way that the low-energy predictions are consistent with the observational data. Some approximate allowed ranges of these free parameters can be determined through correlation analysis allowed by the analytical expressions of the low-energy parameters from the previous section. In addition to the high-energy neutrino parameters, the numerical estimation of the low-energy parameters also depends on two other parameters, namely $I_{\alpha}$ and $\epsilon$. We obtain the values of these two parameters using a numerical approximation method. The detailed methodology of the calculation has been presented in our earlier work \cite{pegu} and we directly present the input values of these two parameters in Table~2. Using these values together with the high-energy input parameters in Eqs.~(\ref{eq:48})--(\ref{eq:50}), (\ref{eq:62})--(\ref{eq:76}) for NO, and Eqs.~(\ref{eq:77})--(\ref{eq:79}) and (\ref{eq:91})--(\ref{eq:100}) for IO, we obtain the corresponding low-energy neutrino parameters. The analysis is performed within the MSSM framework, with the SUSY-breaking scale fixed at $\Lambda_s=1~\mathrm{TeV}$ . We consider three representative values of  $\tan\beta$, namely $10$, $30$, and $50$. In addition to determining the low-energy predictions, we estimate the amount of deviations of the low-energy neutrino parameters from their corresponding high-energy input values and investigate their dependence on  $\tan\beta$. This analysis allows us to examine the effects of  $\tan\beta$ on the low-energy neutrino parameters. The numerical results for the NO and IO scenarios are presented separately in the following subsections.

\begin{table}[t]
\centering
\begin{tabular}{lccc}
\hline
Parameter & $\tan\beta=10$ & $\tan\beta=30$ & $\tan\beta=50$ \\
\hline
$I_\alpha$ & $0.885378$ & $0.884058$  & $0.855389$ \\

$\epsilon$ &$-0.001314$  & $-0.012505$  & $-0.042815$ \\
\hline
\end{tabular}
\caption{Calculated values of $I_\alpha$ and $\epsilon$ for $\tan\beta=10$, $30$, and $50$ with $\Lambda_s=1\,\mathrm{TeV}$.}
\label{tab:Ialpha}
\end{table}
\subsection{Normal Order}
\indent~ In the minimal seesaw framework and NO scenario, we have the high-energy input mass $m_1^{\mu\tau}=0$, and the other two input masses are chosen such that $m_2^{\mu\tau}<m_3^{\mu\tau}$. The appropriate input values of $m_2^{\mu\tau}$,\ $m_3^{\mu\tau}$ and mixing angles $\theta_{13}^{\mu\tau}$,\ $\theta_{12}^{\mu\tau}$ are determined from the correlation analysis allowed by Eqs.(\ref{eq:49}-\ref{eq:50}) and (\ref{eq:62}-\ref{eq:64}). We use the $3\sigma$ ranges of $\Delta m_{21}^2=m_2^2$, $\Delta m_{31}^2=m_3^2$, $\sin^2\theta_{13}$ and $\sin^2\theta_{12}$ from Table~1 in these correlation analyses and estimate the approximate allowed ranges of the high-energy parameters corresponding to these $3\sigma$ ranges. The correlation plots between the low-energy and high-energy masses  ($m_2$ and $m_2^{\mu\tau}$, $m_3$ and $m_3^{\mu\tau}$) and those of  low-energy and high-energy mixing angles ($\sin^2\theta_{13}$ and $\sin\theta_{13}^{\mu\tau}$,  $\sin^2\theta_{12}$ and $\sin\theta_{12}^{\mu\tau}$) for $\tan\beta=10$ and $\Lambda_s=1~\mathrm{TeV}$  are presented in Figs. 1(a)-(d). Correlation plots corresponding to $\tan\beta=30$ and $50$ with $\Lambda_s=1~\mathrm{TeV}$ are presented in Figs.~2(a)-(d), 2(c), 2(d) and 3(a)-(d) respectively. The approximate allowed ranges of $m_{2,\ 3}^{\mu\tau}$ and $\sin\theta_{13,\ 12}^{\mu\tau}$ obtained from these plots are presented in Table~3.

\begin{figure}[t]
\centering

\includegraphics[width=0.46\textwidth]{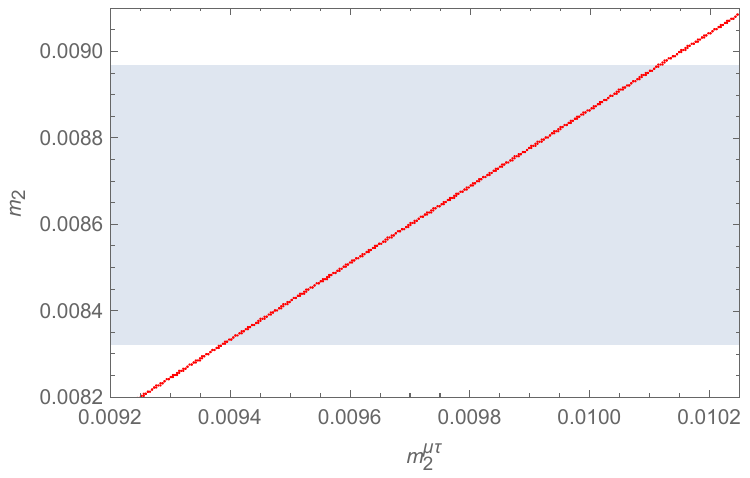}
\hspace{0.02\textwidth}
\includegraphics[width=0.46\textwidth]{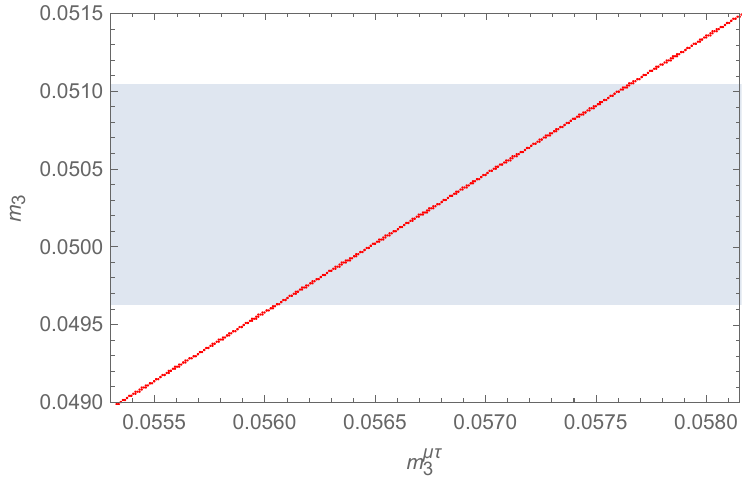}

\vspace{0.1cm}

\makebox[0.46\textwidth]{(a)}
\hspace{0.02\textwidth}
\makebox[0.46\textwidth]{(b)}

\vspace{0.3cm}

\includegraphics[width=0.46\textwidth]{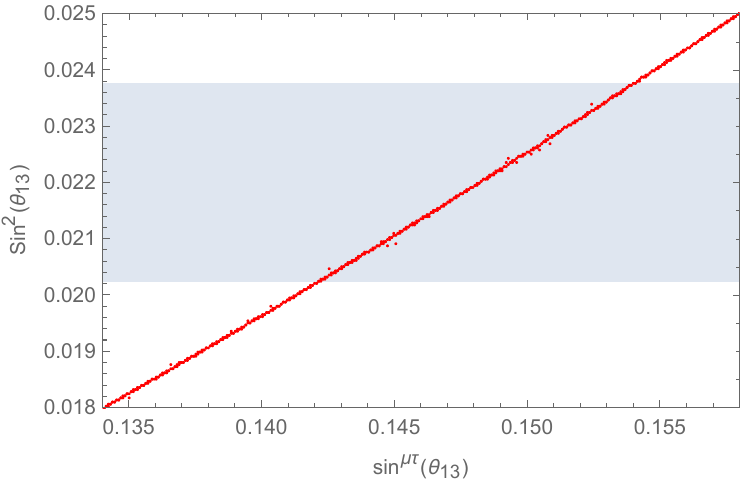}
\hspace{0.02\textwidth}
\includegraphics[width=0.46\textwidth]{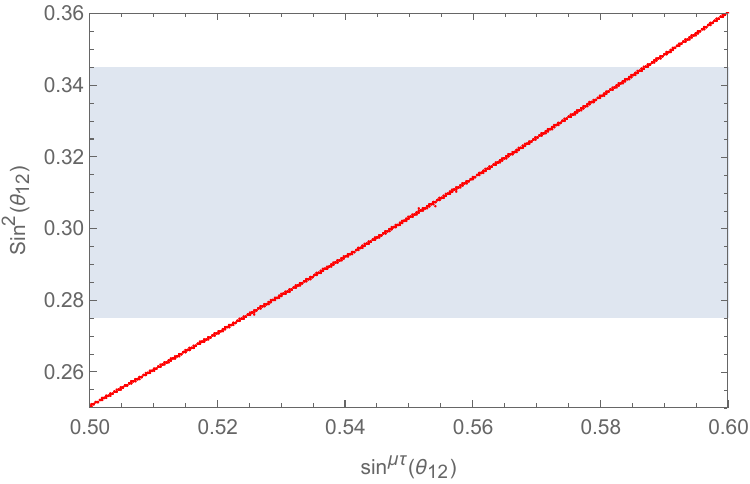}

\vspace{0.1cm}

\makebox[0.46\textwidth]{(c)}
\hspace{0.02\textwidth}
\makebox[0.46\textwidth]{(d)}

\caption{Correlation plots between (a) $m_2$ and $m_2^{\mu\tau}$ (b) $m_3$ and $m_3^{\mu\tau}$, (c) $\sin^2\theta_{13}$ and $\sin^2\theta_{13}^{\mu\tau}$ and (d) $\sin^2\theta_{12}$ and $\sin^2\theta_{12}^{\mu\tau}$ for $\tan\beta=10$ with $\Lambda_s=1\,\mathrm{TeV}$ for the case of NO. The horizontal blue band in each figure represents the $3\sigma$ allowed range of the concerned parameters.}

\label{Fig1}
\end{figure}

\begin{figure}[t]
\centering

\includegraphics[width=0.46\textwidth]{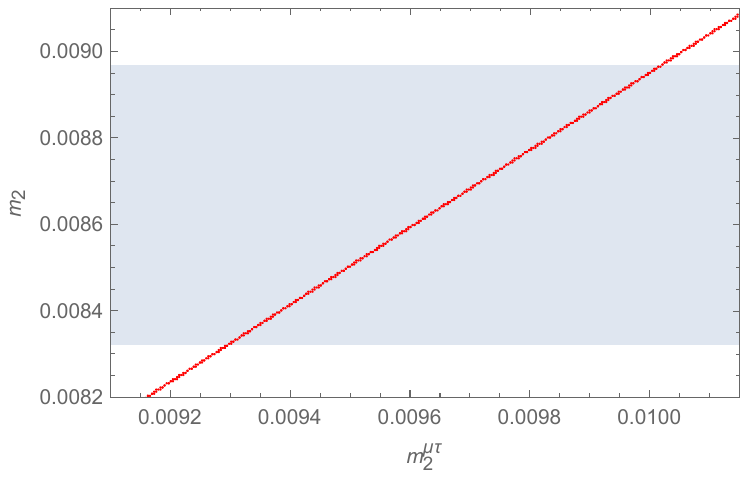}
\hspace{0.02\textwidth}
\includegraphics[width=0.46\textwidth]{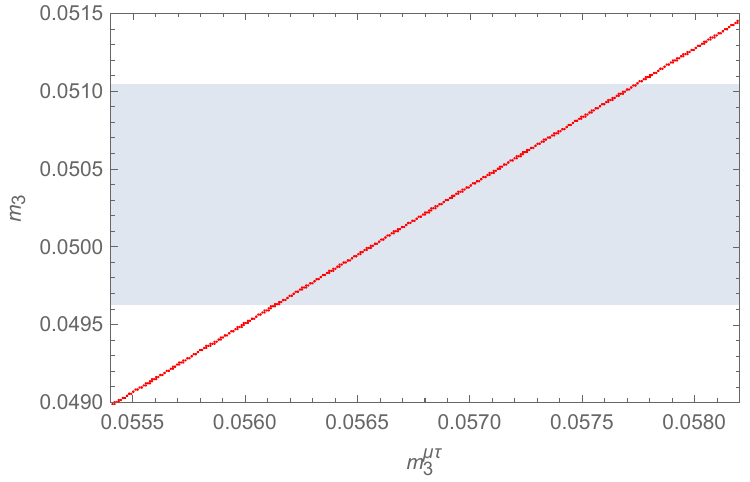}

\vspace{0.1cm}

\makebox[0.46\textwidth]{(a)}
\hspace{0.02\textwidth}
\makebox[0.46\textwidth]{(b)}

\vspace{0.3cm}

\includegraphics[width=0.46\textwidth]{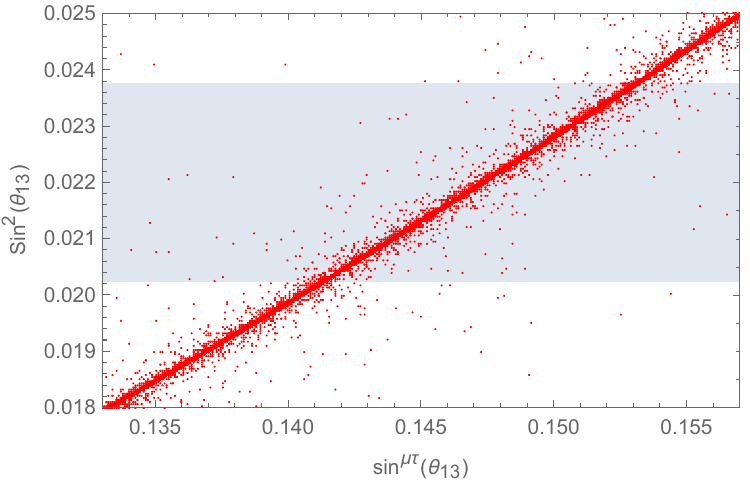}
\hspace{0.02\textwidth}
\includegraphics[width=0.46\textwidth]{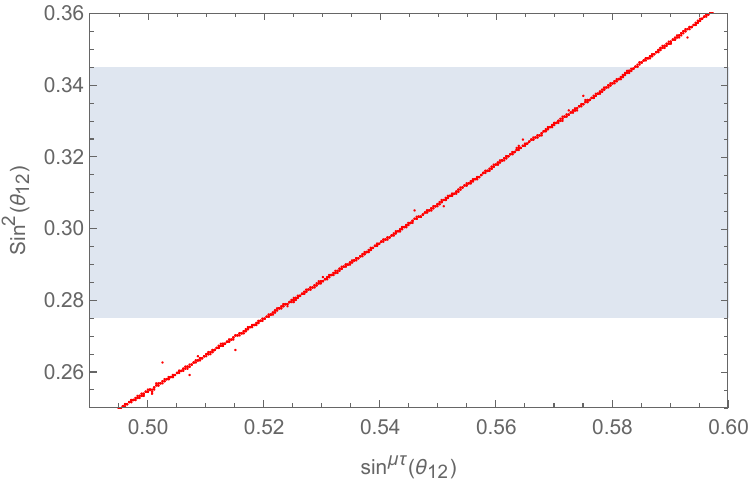}

\vspace{0.1cm}

\makebox[0.46\textwidth]{(c)}
\hspace{0.02\textwidth}
\makebox[0.46\textwidth]{(d)}

\caption{Correlation plots between (a) $m_2$ and $m_2^{\mu\tau}$ (b) $m_3$ and $m_3^{\mu\tau}$, (c) $\sin^2\theta_{13}$ and $\sin^2\theta_{13}^{\mu\tau}$ and (d) $\sin^2\theta_{12}$ and $\sin^2\theta_{12}^{\mu\tau}$ for $\tan\beta=30$ with $\Lambda_s=1\,\mathrm{TeV}$ for the case of NO. The horizontal blue band in each figure represents the $3\sigma$ allowed range of the concerned parameters.}

\label{Fig2}
\end{figure}
\begin{figure}[t]
\centering

\includegraphics[width=0.46\textwidth]{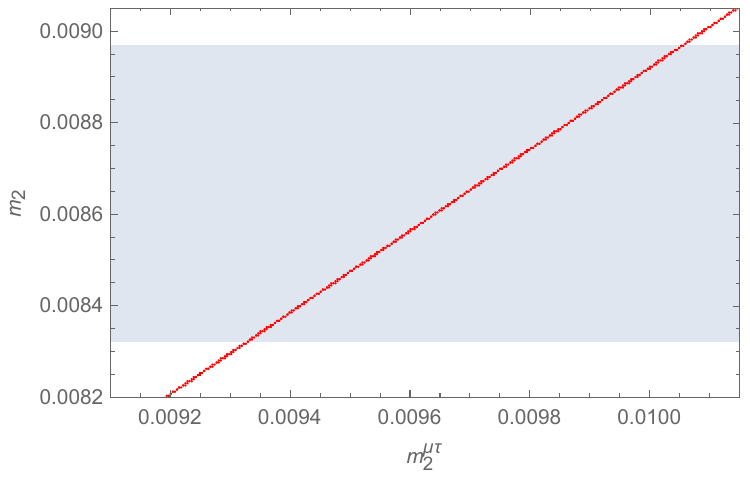}
\hspace{0.02\textwidth}
\includegraphics[width=0.46\textwidth]{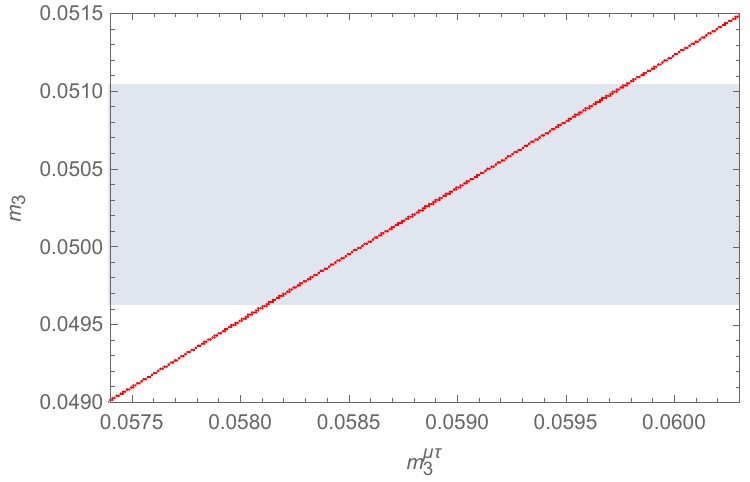}

\vspace{0.1cm}

\makebox[0.46\textwidth]{(a)}
\hspace{0.02\textwidth}
\makebox[0.46\textwidth]{(b)}

\vspace{0.3cm}

\includegraphics[width=0.46\textwidth]{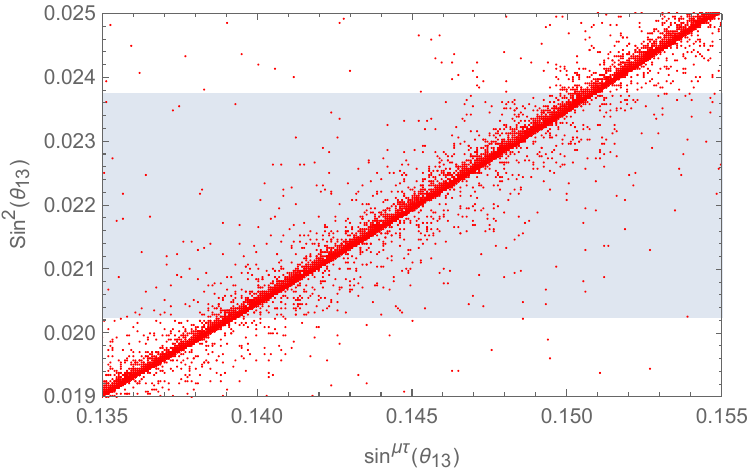}
\hspace{0.02\textwidth}
\includegraphics[width=0.46\textwidth]{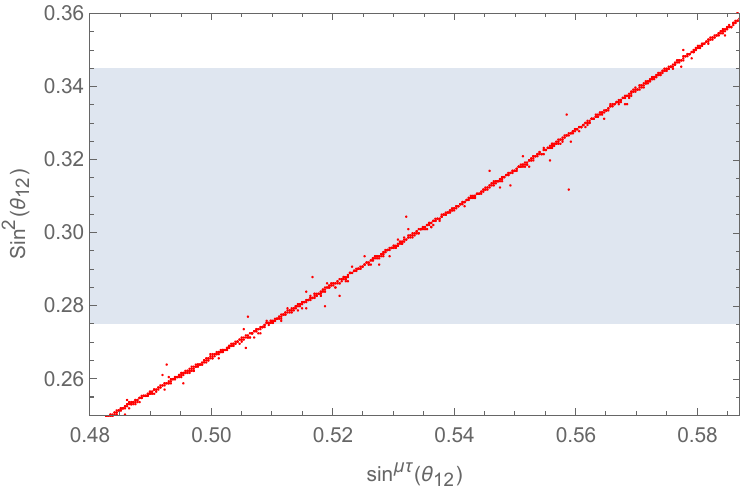}

\vspace{0.1cm}

\makebox[0.46\textwidth]{(c)}
\hspace{0.02\textwidth}
\makebox[0.46\textwidth]{(d)}

\caption{Correlation plots between (a) $m_2$ and $m_2^{\mu\tau}$ (b) $m_3$ and $m_3^{\mu\tau}$, (c) $\sin^2\theta_{13}$ and $\sin^2\theta_{13}^{\mu\tau}$ and (d) $\sin^2\theta_{12}$ and $\sin^2\theta_{12}^{\mu\tau}$ for $\tan\beta=50$ with $\Lambda_s=1\,\mathrm{TeV}$ for the case of NO. The horizontal blue band in each figure represents the $3\sigma$ allowed range of the concerned parameters.}
\label{Fig3}
\end{figure}

\indent~ We now choose appropriate input values for $m_{2,\ 3}^{\mu\tau}$, $\sin\theta_{13}^{\mu\tau}$ and $\sin\theta_{12}^{\mu\tau}$ from the allowed ranges given in Table 3. Besides these free parameters, the input values of $\theta_{23}^{\mu\tau}$ and the CP phases are fixed by the reflection symmetry at maximal values. Since there exists two possible scenarios, viz.,  Case-I ($\delta^{\mu\tau}=\pi/2$) and Case-II ($\delta^{\mu\tau}=3\pi/2$), corresponding high-energy input values of CP phases are taken as per Eqs.~(\ref{eq:14}) and (\ref{eq:15}). Finally, using all the input values of high-energy free parameters in Eqs.(\ref{eq:49}-\ref{eq:50}), (\ref{eq:62}-\ref{eq:64}) and (\ref{eq:65}-\ref{eq:77}), we estimate the low-energy values of all the parameters at $\Lambda_{EW}$ for $\tan\beta=10,\ 30$ and $\tan\beta=50$ with $\Lambda_s=1\ TeV$. The high energy input values and corresponding low energy predictions are presented in Table 4. The table consists of three horizontal blocks, where the first block represents the values of mass eigenvalues and mixing angles, second and third blocks represent the values of the CP phases, Jarlskog invariant and the effective Majorana neutrino mass for Case-I and Case-II respectively. Table 4 also comprises of four vertical blocks. The first block represents the high-energy input values of all the parameters. The second block represents the corresponding low-energy predictions for $\tan\beta=10$. The third and fourth vertical blocks represent the predictions for $\tan\beta=30$ and $50$ respectively. As for example, for $\tan\beta=10$ we choose the input values $m_2^{\mu\tau}= 0.0096\ eV$ and $m_3^{\mu\tau}=0.0565\ eV$, such that we obtain low-energy masses as $m_2=0.008773\ eV$ and $m_3=0.050462\ eV$ at $\Lambda_{EW}$. From Table 1, we see that the resulting mass-squared differences $\Delta m^2_{21}=7.69\times 10^{-5}\ eV^2$ and $\Delta m^2_{31}=2.55\times 10^{-3}\ eV^2$ are close to their respective global best-fit values. The predicted sum of mass eigenvalues, $\sum_i m_i = 0.0592\ \mathrm{eV}$, also lies well below
 the cosmological upper bound $0.12\ \mathrm{eV}$.  Further, with the input values $\sin^2\theta_{12}^{\mu\tau}=0.3025$, $\sin^2\theta_{23}^{\mu\tau}=0.5$ and $\sin^2\theta_{13}^{\mu\tau}=0.0222$, we obtain the corresponding low-energy values $\sin^2\theta_{12}=0.302534$, $\sin^2\theta_{23}=0.50003$ and $\sin^2\theta_{13}=0.02224$. We see that the predicted low-energy values of $\sin^2\theta_{12}$ and $\sin^2\theta_{13}$ are close to their respective experimental best-fit values of $0.307$ (without SK data) and $0.02215$ (with SK data) respectively. The low-energy predicted value of $\sin^2\theta_{23}=0.50003$ shows a very small deviation from the maximal value $\sin^2\theta_{23}=0.5$. Nonetheless, the prediction is consistent with the $3\sigma$ range of global data. It also indicates a tendency of $\theta_{23}$ to lie in the second octant under radiative corrections. We next obtain the low-energy values of the CP phases, the Jarlskog invariant, and the effective Majorana neutrino mass from Eqs.~(\ref{eq:65})--(\ref{eq:77}) for $\tan\beta=10$. The high-energy input values and the corresponding low-energy predictions for Case-I and Case-II are presented in the second and third horizontal blocks of Table~4, respectively. For Case-I, we obtain $\delta = 90.0002^\circ$, $\sigma = 270^\circ$, $\phi_1 = 90.0001^\circ$ and $\phi_2 = \phi_3 = 0^\circ$. For Case-II, the corresponding values are $\delta = 269.999^\circ$, $\sigma = 90.0014^\circ$, $\phi_1 = 270^\circ$ and $\phi_2 = \phi_3 = 0^\circ$. We observe that these CP phases receive very negligible contributions from radiative corrections. Further, we find that the low-energy prediction of $J$ is $0.0331$ for Case-I and $-0.0331$ for Case-II. These predictions are consistent with the experimental bounds $J=0.0333$. Moreover, we obtain $|m_{ee}| = 0.00372\ \mathrm{eV}$ for both Case-I and Case-II. This predicted value lies well within the most stringent upper bound provided by the KamLAND-Zen Collaboration, which reports $|\langle m \rangle_{ee}| < (0.028 - 0.122) \mathrm{eV}$ \cite{Kam}.\\
\indent~ Next, keeping the same set of high-energy input values as those used for $\tan\beta=10$, we calculate the low-energy values of all parameters for  $\tan\beta=30$ and $50$. The corresponding low-energy predictions are presented in the third and fourth vertical block of Table 4 respectively. For the case of $\tan\beta=30\ (50)$, we obtain $m_2 = 0.008599(0.008394)\ \mathrm{eV}$ and $m_3 = 0.050243 (0.049967)\ \mathrm{eV}$. We see that, for both cases, the predicted mass eigenvalues yield mass-squared differences that are in good agreement with their respective experimental best-fit values.  The predicted mass eigenvalues are also consistent with the cosmological upper bound, $\sum_i m_i < 0.12 \mathrm{eV}$. Further, for $\tan\beta = 30(50)$, we obtain $\sin^2\theta_{13} = 0.022420(0.023210)$, $\sin^2\theta_{23} = 0.50029(0.50081)$ and $\sin^2\theta_{12} = 0.302730(0.306780)$. Similar to the case of $\tan\beta=10$, we observe that the predicted low-energy values of $\sin^2\theta_{13}$ and $\sin^2\theta_{12}$ are close to their respective experimental best-fit values of $0.02215$ (with SK data) and $0.307$ (without SK data) respectively. Moreover, the predicted value of $\sin^2\theta_{23}$ lies within the corresponding experimental $3\sigma$ ranges, with low-energy values $0.5003$ and $0.5008$ at $\tan\beta=30$ and $50$ respectively. Turning to the low-energy predictions for the CP phases, for $\tan\beta=30\ (50)$ we obtain $\delta=90.002^\circ\ (90.004^\circ)$ for Case-I, while for Case-II we obtain $\delta=269.998^\circ\ (269.996^\circ)$. Similar results are also obtained for other CP phases as can be read off Table 4. Low-energy values of $J$ and $|m_{ee}|$ are found to be similar as those obtained for $\tan\beta=10$. All these predictions are consistent with experimental data. One concluding remark is that the same set of high-energy input values employed for $\tan\beta=10$ also leads to consistent low-energy predictions for $\tan\beta=30$ and $50$.
\begin{table}[t]
\centering
\begin{tabular}{l c|c|c}
\hline
&$\tan\beta=10$ & $\tan\beta=30$ &$\tan\beta=50$ \\
\hline
Parameter & $1\ \mathrm{TeV}$& $1\ \mathrm{TeV}$  & $1\ \mathrm{TeV}$  \\
 \hline
$m_2^{\mu\tau}$ & 0-0.010117 & 0-0.010024 & 0-0.010055 \\
$m_3^{\mu\tau}$ & 0-0.057685& 0-0.057780 & 0-0.059788 \\
$\sin\theta_{13}^{\mu\tau}$ & 0.141604-0.154119& 0.140842-0.153673 & 0.138748-0.151422 \\
$\sin\theta_{12}^{\mu\tau}$ & 0.523802-0.586942& 0.518727-0.584545 & 0.507702-0.577650 \\
\hline
\end{tabular}\caption{ Approximate allowed ranges of $m_i^{\mu\tau}$, $\sin\theta_{12}^{\mu\tau}$ and $\sin\theta_{13}^{\mu\tau}$ for $\tan\beta= 10,\ 30$ and $50$  with $\Lambda_s=  1\ TeV$ for the case of NO.}
\end{table} 

\begin{table}[H]
\centering
\begin{tabular}{lc|c|c|c}
\hline
& High energy value & \multicolumn{2}{c}{Low energy value at $\Lambda_{EW}$} \\
\hline
& & $\tan\beta=10$& $\tan\beta=30$ & $\tan\beta=50$ \\
\hline
Parameter & $10^{14}\ \mathrm{GeV}$ &$1\ \mathrm{TeV}$& $1\ \mathrm{TeV}$ & $1\ \mathrm{TeV}$ \\
\hline
$m_1\left(eV\right)$ &0&0&0&0\\
$m_2\left(eV\right)$ & 0.0096 & 0.008773 & 0.008599 & 0.008394 \\
$m_3\left(eV\right)$ & 0.0565 &0.050462& 0.050243 & 0.049967  \\
$\sin^2\theta_{13}$ & 0.0222 &0.022240& 0.022420 & 0.023210 \\
$\sin^2\theta_{23}$ & 0.5 &0.50003& 0.50029 & 0.50081 \\
$\sin^2\theta_{12}$ & 0.3025 & 0.302534& 0.302730 &  0.306780 \\
$\Delta m_{21}^2=m_2^2(/10^{-5})\ eV^2$& -& 7.69& 7.39& 7.05\\
$\Delta m_{31}^2=m_3^2(/10^{-3})\ eV^2$& -& 2.55 & 2.52 & 2.49\\
$\Sigma_i m_i \left(eV\right)$ & - &0.0592& 0.0587 & 0.05507 \\
 \hline
Case-I \\ \hline
$\delta\left(/^\circ\right)$ & 90 & 90.0002 & 90.002&90.004\\
$\sigma\left(/^\circ\right)$ & 270 & 269.999& 269.998 & 269.996\\
$\phi_1\left(/^\circ\right)$ & 90 & 90.0001 & 90.001& 90.003\\
$\phi_2\left(/^\circ\right)$ & 0 & 0 & 0&0 \\
$\phi_3\left(/^\circ\right)$ & 0 & 0 & 0&0\\
$J$&- & 0.0331 & 0.0332&0.0333 \\
$|\langle m \rangle_{ee}|$&- & 0.00372 &0.00374 & 0.00389\\
\hline
Case-II \\ \hline
$\delta\left(/^\circ\right)$ & 270 & 269.999& 269.998  & 269.996  \\
$\sigma\left(/^\circ\right)$ & 90 &90.0014& 90.002 & 90.004 \\
$\phi_1\left(/^\circ\right)$ & 270 &269.999& 269.999 & 269.997 \\
$\phi_2\left(/^\circ\right)$ & 0 & 0 & 0&0  \\
$\phi_3\left(/^\circ\right)$ & 0 & 0 & 0&0 \\
$J$&- & -0.0331 & -0.0332&-0.0333 \\
$|\langle m \rangle_{ee}|$&- & 0.00372 &0.00374 & 0.00389\\
\hline
\end{tabular}
\caption{High energy values of parameters at $\Lambda_{\mu\tau}=10^{14} \ GeV$ and corresponding low energy values at $m_t$ scale for $\Lambda_s=1 \ TeV$ with $\tan\beta= 10,\ 30$ and $50$  for the case of NO.}
\end{table}
\begin{table}[H]
\centering
\begin{tabular}{lccc}
\hline
Parameter & \multicolumn{3}{c}{Amount of deviations at $\Lambda_{EW}$ }\\
\cline{2-4}
& $\tan\beta=10$ & $\tan\beta=30$ & $\tan\beta=50$\\
\hline
$\Delta m_1~(\mathrm{eV})$ &0&0&0\\
$\Delta m_2(\mathrm{eV})$ &0.000827&0.001001&0.001206\\
$\Delta m_3(\mathrm{eV})$ &0.006038&0.006257&0.006533\\
$\Delta(\sin^2\theta_{13})$ &0.00004&0.00027&0.00101\\
$\Delta(\sin^2\theta_{23})$ &0.00003&0.00029&0.00081\\
$\Delta(\sin^2\theta_{12})$ &0.00003&0.00023&0.00428\\
\hline
\multicolumn{4}{c}{\textbf{Case-I}}\\
\hline
$\Delta\delta~(^\circ)$ &0.0002&0.002&0.004\\
$\Delta\sigma~(^\circ)$ &0.001&0.002&0.004\\
$\Delta\phi_1~(^\circ)$ &0.0001&0.001&0.003\\
$\Delta\phi_2~(^\circ)$ &0&0&0\\
$\Delta\phi_3~(^\circ)$ &0&0&0\\
\hline
\multicolumn{4}{c}{\textbf{Case-II}}\\
\hline
$\Delta\delta~(^\circ)$ &0.001&0.002&0.004\\
$\Delta\sigma~(^\circ)$ &0.0014&0.002&0.004\\
$\Delta\phi_1~(^\circ)$ &0.001&0.001&0.003\\
$\Delta\phi_2~(^\circ)$ &0&0&0\\
$\Delta\phi_3~(^\circ)$ &0&0&0\\
\hline
\end{tabular}

\caption{Amount of deviations of parameters at $\Lambda_{EW}$ scale with respect to their high-energy input values for $\Lambda_s=1 \ TeV$ with $\tan\beta= 10,30, 50$ for the case of NO.}
\end{table}

 \indent~ We estimate the amount of deviations of the neutrino parameters at $\Lambda_{EW}$ from their corresponding high-energy input values and see how these deviations vary with the variations of $\tan\beta$. We define the amount of deviations in the mass eigenvalues as $\Delta m_i= (m_i^{\mu\tau}-m_i)$ $(i=2, 3)$, in the mixing angles as $\Delta (\sin^2\theta_{ij})=(\sin^2\theta_{ij}^{\mu\tau}-\sin^2\theta_{ij})$ $(ij=12, 23, 13)$ and in the CP phases as $\Delta x=(x^{\mu\tau}-x)$ $(x=\delta,\ \sigma,\ \phi_1,\ \phi_2,\ \phi_3)$. The resulting deviations in the mass eigenvalues, mixing angles and CP phases are given in Table~5. The second, third, and fourth vertical blocks of Table~5 correspond to the amount of deviations for $\tan\beta=10$, $30$ and $50$ respectively. It is apparent that both $\Delta m_2$ and $\Delta m_3$ increase with increasing $\tan\beta$. This can be realized analytically from the Eqs.~(\ref{eq:49})--(\ref{eq:50}). As for example, low-energy $m_3$ (Eq.(\ref{eq:50})) depends on two factors : $I_{\alpha}$ and $1-|\epsilon| \left( c_{13}^{\mu\tau}\right)^2 $. From Table 2, we see that $I_{\alpha}$ decreases with the increase of $\tan\beta$. On the other hand $|\epsilon|$ increases, resulting a decrease of the factor $1-|\epsilon| \left( c_{13}^{\mu\tau}\right)^2 $ with increase of $\tan\beta$. Thus, both $I_{\alpha}$ and $1-|\epsilon| \left( c_{13}^{\mu\tau}\right)^2 $ decrease simultaneously with increasing $\tan\beta$ resulting into a lower and lower value of $m_3$. The same behavior can also be understood for $\Delta m_2$ from Eq.~(\ref{eq:49}). A similar enhancement with increasing $\tan\beta$ is observed in the deviations of the mixing angles. The deviations $\Delta(\sin^2\theta_{13})$, $\Delta(\sin^2\theta_{23})$ and $\Delta(\sin^2\theta_{12})$ all increase with $\tan\beta$. This behavior follows from Eqs.~(\ref{eq:62})--(\ref{eq:64}), in which the first term on the right-hand side corresponds to the high-energy input value, while the second term represents the radiative correction involving $\epsilon$ and the high-energy masses and mixing angles. For example, the expression for $\sin^2\theta_{13}$ in Eq.~(\ref{eq:62}) consists of the high-energy contribution $(s_{13}^{\mu\tau})^2$ and a correction proportional to $|\epsilon|(s_{13}^{\mu\tau})^2c_{13}^{\mu\tau}\left[(c_{12}^{\mu\tau})^2+(s_{12}^{\mu\tau})^2\Delta_{32}\right]$. For the chosen input values of the neutrino masses and mixing angles, the correction term $|\epsilon|(s_{13}^{\mu\tau})^2c_{13}^{\mu\tau}\left[(c_{12}^{\mu\tau})^2+(s_{12}^{\mu\tau})^2\Delta_{32}\right]$ is positive and increases with increasing $|\epsilon|$. Since the high-energy input values of the masses and mixing angles are kept fixed for $\tan\beta=10$, $30$, and $50$, the variation of the radiative correction is governed primarily by the change in $\epsilon$. Therefore, the increase in $|\epsilon|$ with $\tan\beta$ leads to a progressively larger value of $\sin^2\theta_{13}$ and, consequently, to an increasing $\Delta(\sin^2\theta_{13})$. The same reasoning applies to Eqs.~(\ref{eq:63})--(\ref{eq:64}), resulting in the corresponding increase in $\Delta(\sin^2\theta_{23})$ and $\Delta(\sin^2\theta_{12})$. The estimated deviations $\Delta\delta$ and $\Delta\sigma$ are given in Table~5. Both deviations increase with increasing $\tan\beta$. From Eqs.~(\ref{eq:71}) and (\ref{eq:73}), the deviations are primarily controlled by $\epsilon$ and the ratios  $(J_1^{\mu\tau}+ J_2^{\mu\tau})/J_0^{\mu\tau}$ and $ (\langle m \rangle_{ee})_1^{\mu\tau}/ (\langle m \rangle_{ee})_0^{\mu\tau}$. For the chosen input values in Table~4, these ratios are approximately 
  $(J_1^{\mu\tau}+ J_2^{\mu\tau})/J_0^{\mu\tau}\sim 0.053$ and $ (\langle m \rangle_{ee})_1^{\mu\tau}/ (\langle m \rangle_{ee})_0^{\mu\tau}\sim -0.05$. These relatively small ratios, together with the $|\epsilon|$, suppress the radiative corrections to the CP phases, resulting in only small deviations of $\delta$ and $\sigma$ from their high-energy input values. Nevertheless, $|\epsilon|$ increases with increasing $\tan\beta$, consequently, both $\Delta\delta$ and $\Delta\sigma$ exhibit an increasing trend with $\tan\beta$. A similar argument applies to the unphysical phase $\phi_1$. From Eq.~(\ref{eq:74}), the first and second terms provide the dominant contributions to the predicted low-energy value of $\phi_1$, whereas the third term gives a negligible contribution. This suppression arises from the small factors $\epsilon/2$ and $s_{13}^{\mu\tau}$ appearing in the corresponding numerator, while the denominator is of order unity. Consequently, the third term has a negligible effect on $\phi_1$. The deviation of $\phi_1$ from its high-energy input value is therefore determined predominantly by the second term, which is proportional to $\epsilon(J_1^{\mu\tau}+J_2^{\mu\tau})/J_0^{\mu\tau}$. A similar suppression occurs for the unphysical phases $\phi_2$ and $\phi_3$. In Eqs.~(\ref{eq:75}) and (\ref{eq:76}), the terms contributing to their deviations contain analogous suppression factors to those appearing in the third term of Eq.~(\ref{eq:74}). These terms are therefore negligible, resulting in only very small deviations of $\phi_2$ and $\phi_3$ from their corresponding high-energy input values.

 \subsection{Inverted Order ($m_2>m_1$)}
 \indent~ We perform the numerical analysis for the IO scenario following the same procedure as that adopted for the NO scenario. In the IO scenario, we take $m_3^{\mu\tau}=0$ with $m_2^{\mu\tau}>m_1^{\mu\tau}$. The high-energy input values of $m_{1,\ 2}^{\mu\tau}$ and $\sin\theta_{12, \ 13}^{\mu\tau}$ are determined from the correlation analyses based on Eqs.~(\ref{eq:78})--(\ref{eq:79}) and (\ref{eq:91})--(\ref{eq:93}). We use the experimentally allowed $3\sigma$ ranges of $|\Delta m_{31}^2|$, $\Delta m_{21}^2$, $\sin^2\theta_{12}$ and $\sin^2\theta_{13}$ given in Table~1 to obtain the approximate allowed ranges of the high-energy input parameters. The resulting correlation plots for $\tan\beta=10$, $30$ and $50$ with $\Lambda_s=1~\mathrm{TeV}$ are shown in Figs.~4(a)--(d), 5(a)--(d), and 6(a)--(d), respectively. The corresponding approximate allowed ranges of $m_{1,2}^{\mu\tau}$ and $\sin\theta_{12, \ 13}^{\mu\tau}$ are presented in Table~6.\\
 \indent~ We then choose appropriate values of $m_{1,2}^{\mu\tau}$ and $\sin\theta_{12, \ 13}^{\mu\tau}$ from the allowed ranges in Table~6. The high-energy value of $\theta_{23}^{\mu\tau}$, along with the CP phases for Case-I and Case-II, are fixed by the $\mu$--$\tau$ reflection symmetry according to Eqs.~(\ref{eq:14}) and (\ref{eq:15}), respectively. Substituting all high-energy input values into the analytical expressions for the neutrino masses, mixing angles and CP phases, we obtain the corresponding low-energy predictions at $\Lambda_{EW}$ for $\tan\beta=10$, $30$ and $50$, with $\Lambda_s=1~\mathrm{TeV}$.
 \begin{figure}[t]
\centering

\includegraphics[width=0.46\textwidth]{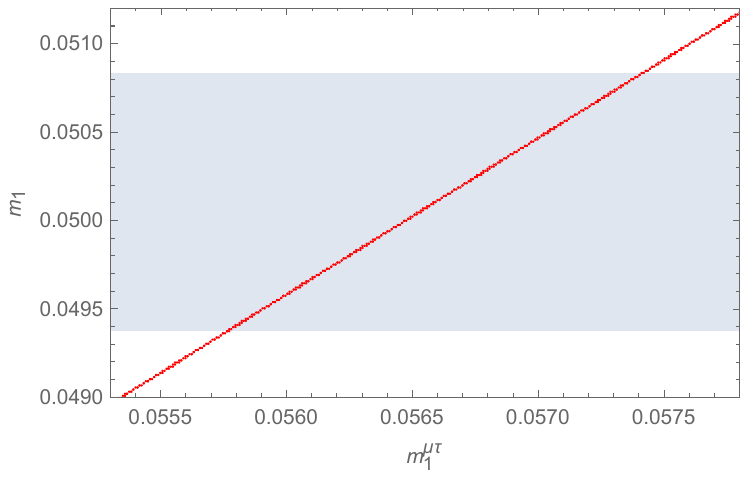}
\hspace{0.02\textwidth}
\includegraphics[width=0.46\textwidth]{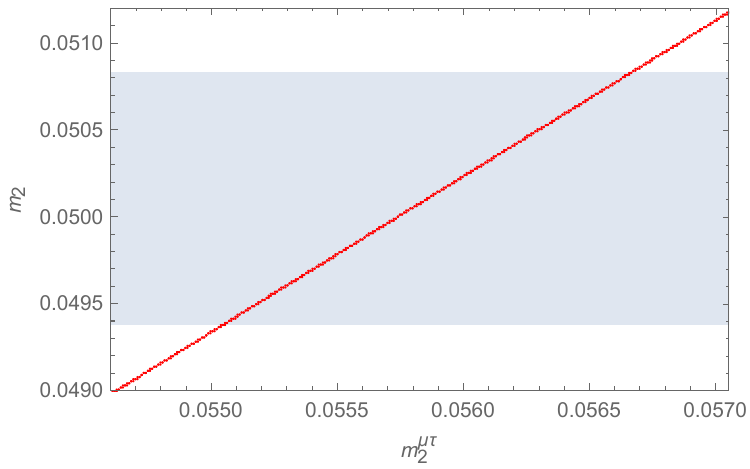}

\vspace{0.1cm}

\makebox[0.46\textwidth]{(a)}
\hspace{0.02\textwidth}
\makebox[0.46\textwidth]{(b)}

\vspace{0.3cm}

\includegraphics[width=0.46\textwidth]{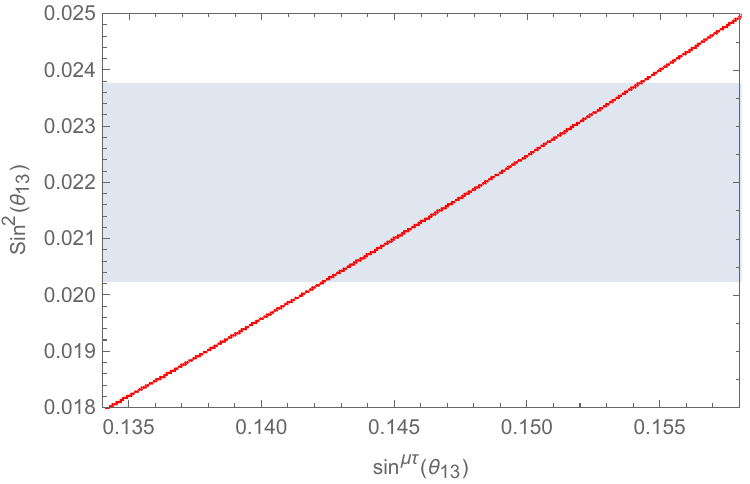}
\hspace{0.02\textwidth}
\includegraphics[width=0.46\textwidth]{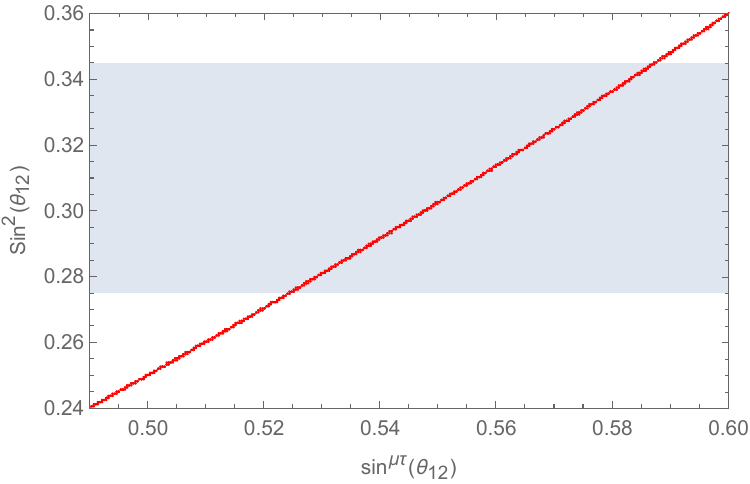}

\vspace{0.1cm}

\makebox[0.46\textwidth]{(c)}
\hspace{0.02\textwidth}
\makebox[0.46\textwidth]{(d)}

\caption{Correlation plots between (a) $m_1$ and $m_1^{\mu\tau}$ (b) $m_2$ and $m_2^{\mu\tau}$, (c) $\sin^2\theta_{13}$ and $\sin^2\theta_{13}^{\mu\tau}$ and (d) $\sin^2\theta_{12}$ and $\sin^2\theta_{12}^{\mu\tau}$ for $\tan\beta=10$ with $\Lambda_s=1\,\mathrm{TeV}$ for the case of IO. The horizontal blue band in each figure represents the $3\sigma$ allowed range of the concerned parameters.}

\label{Fig4}
\end{figure}
\begin{figure}[t]
\centering

\includegraphics[width=0.46\textwidth]{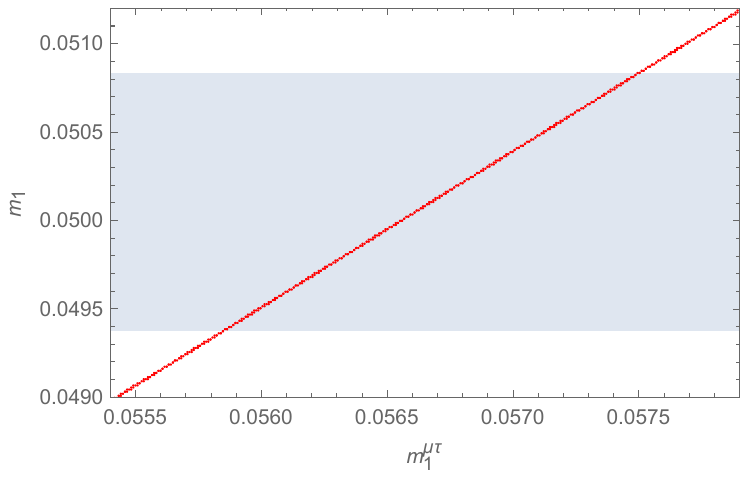}
\hspace{0.02\textwidth}
\includegraphics[width=0.46\textwidth]{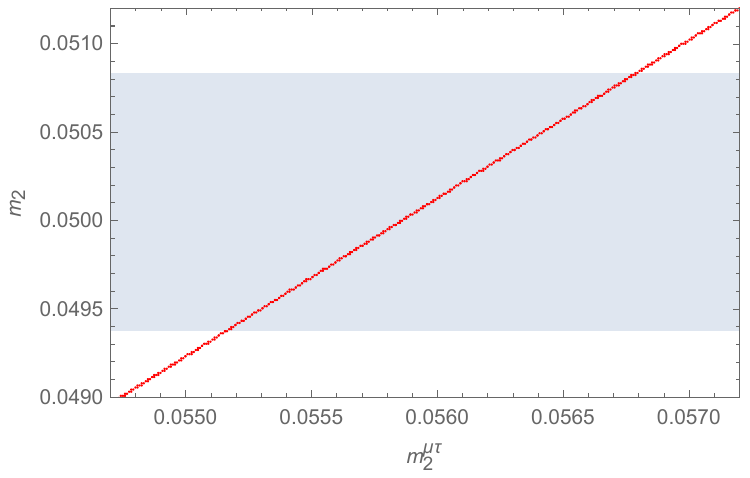}

\vspace{0.1cm}

\makebox[0.46\textwidth]{(a)}
\hspace{0.02\textwidth}
\makebox[0.46\textwidth]{(b)}

\vspace{0.3cm}

\includegraphics[width=0.46\textwidth]{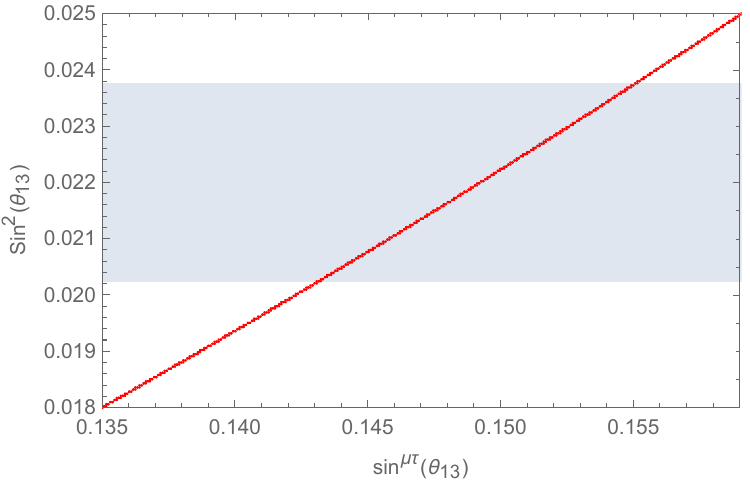}
\hspace{0.02\textwidth}
\includegraphics[width=0.46\textwidth]{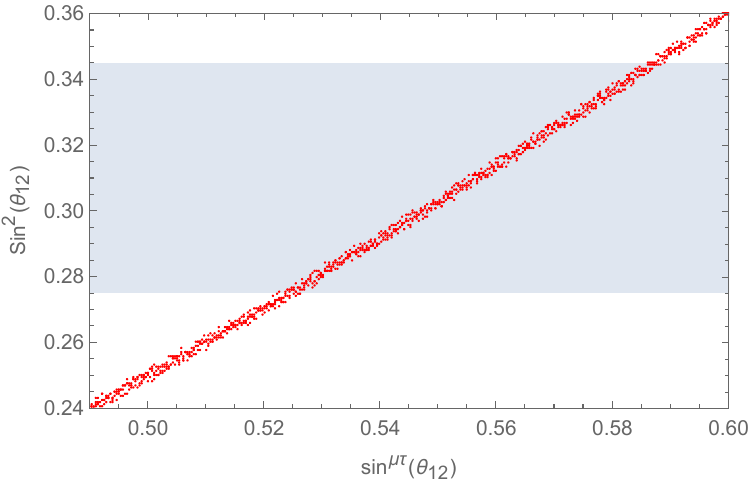}

\vspace{0.1cm}

\makebox[0.46\textwidth]{(c)}
\hspace{0.02\textwidth}
\makebox[0.46\textwidth]{(d)}

\caption{Correlation plots between (a) $m_1$ and $m_1^{\mu\tau}$ (b) $m_2$ and $m_2^{\mu\tau}$, (c) $\sin^2\theta_{13}$ and $\sin^2\theta_{13}^{\mu\tau}$ and (d) $\sin^2\theta_{12}$ and $\sin^2\theta_{12}^{\mu\tau}$ for $\tan\beta=30$ with $\Lambda_s=1\,\mathrm{TeV}$ for the case of IO. The horizontal blue band in each figure represents the $3\sigma$ allowed range of the concerned parameters.}

\label{Fig5}
\end{figure}

\begin{figure}[t]
\centering

\includegraphics[width=0.46\textwidth]{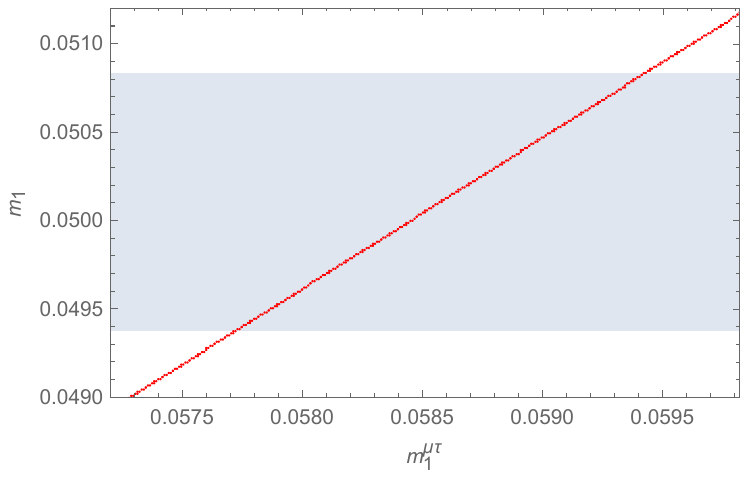}
\hspace{0.02\textwidth}
\includegraphics[width=0.46\textwidth]{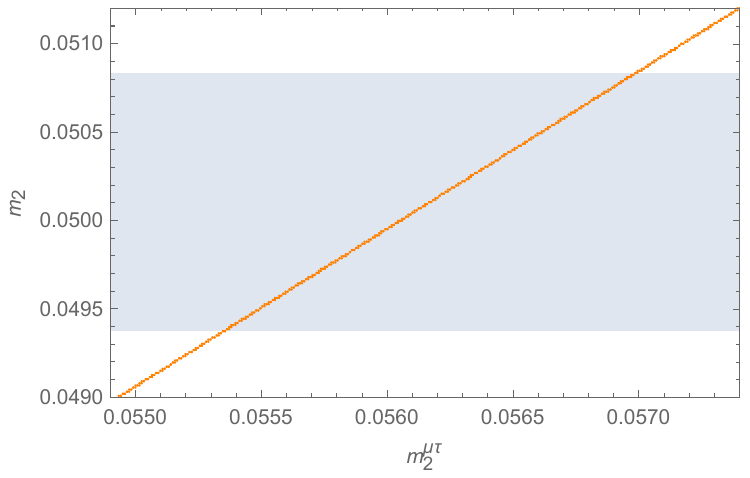}

\vspace{0.1cm}

\makebox[0.46\textwidth]{(a)}
\hspace{0.02\textwidth}
\makebox[0.46\textwidth]{(b)}

\vspace{0.3cm}

\includegraphics[width=0.46\textwidth]{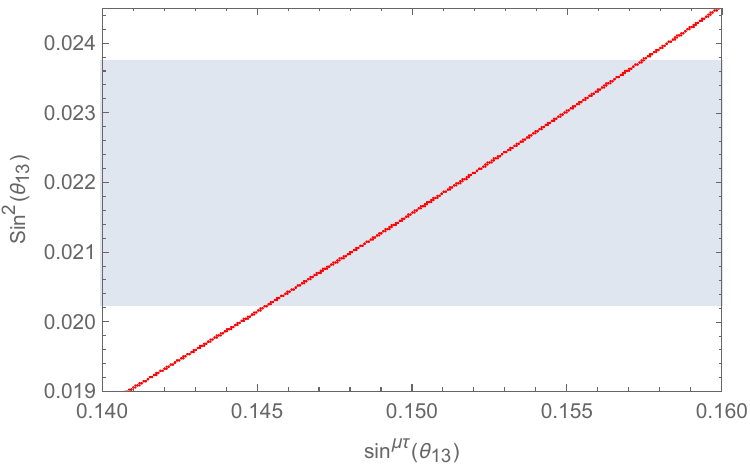}
\hspace{0.02\textwidth}
\includegraphics[width=0.46\textwidth]{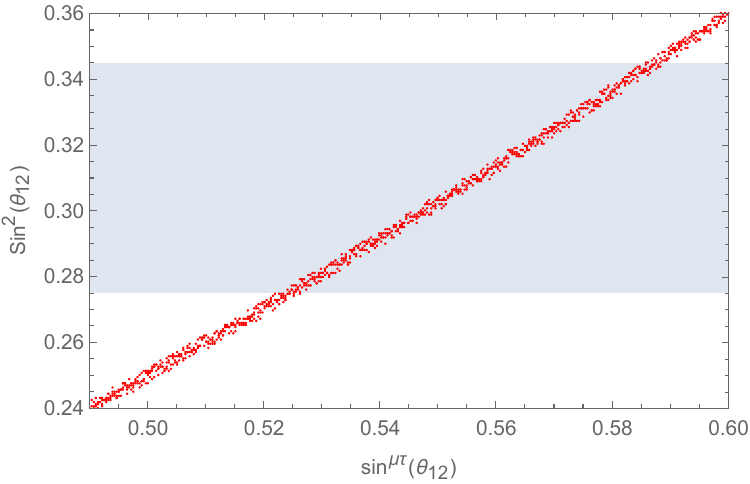}

\vspace{0.1cm}

\makebox[0.46\textwidth]{(c)}
\hspace{0.02\textwidth}
\makebox[0.46\textwidth]{(d)}

\caption{Correlation plots between (a) $m_1$ and $m_1^{\mu\tau}$ (b) $m_2$ and $m_2^{\mu\tau}$, (c) $\sin^2\theta_{13}$ and $\sin^2\theta_{13}^{\mu\tau}$ and (d) $\sin^2\theta_{12}$ and $\sin^2\theta_{12}^{\mu\tau}$ for $\tan\beta=50$ with $\Lambda_s=1\,\mathrm{TeV}$ for the case of IO. The horizontal blue band in each figure represents the $3\sigma$ allowed range of the concerned parameters.}
\label{Fig6}
\end{figure}
 The numerical results for the IO scenario are presented in Table~7. The structure of Table~7 is same as that of Table~4. The first horizontal block contains the mass eigenvalues and mixing angles, while the second and third horizontal blocks contain the CP phases, the Jarlskog invariant $J$ and the effective Majorana neutrino mass $|m_{ee}|$ for Case-I and Case-II, respectively. The vertical blocks contain the high-energy input values and the corresponding low-energy predictions for $\tan\beta=10$, $30$ and $50$.\\
 \indent~ For $\tan\beta=10$, we choose the high-energy input values $m_1^{\mu\tau}=0.054~\mathrm{eV}$ and $m_2^{\mu\tau}=0.055~\mathrm{eV}$. The corresponding low-energy masses at $\Lambda_{EW}$ are $m_1=0.050482~\mathrm{eV}$ and $m_2=0.051255~\mathrm{eV}$. Thus, the sum of the neutrino masses is $\sum_i m_i=0.101737~\mathrm{eV}$, which is below the cosmological upper bound $\sum_i m_i<0.12~\mathrm{eV}$. The resulting mass-squared differences are $\Delta m_{21}^2=7.86\times10^{-5}~\mathrm{eV}^2$ and $|\Delta m_{31}^2|=2.54\times10^{-3}~\mathrm{eV}^2$, both of which are compatible with the corresponding $3\sigma$ ranges and close to the global best-fit values. For the mixing angles, the low-energy predictions are $\sin^2\theta_{13}=0.02219$, $\sin^2\theta_{23}=0.49999$, and $\sin^2\theta_{12}=0.30811$. The predicted values of $\sin^2\theta_{13}$ and $\sin^2\theta_{12}$ are close to their respective experimental best-fit values, while $\sin^2\theta_{23}$ remains within its allowed $3\sigma$ range. The predicted value $\sin^2\theta_{23}<0.5$ indicates a slight preference for the first octant, $\theta_{23}<45^\circ$, in the IO scenario. This is opposite to the tendency observed in the NO case, where the radiatively corrected value lies slightly above the maximal value. The corresponding low-energy CP phases for Case-I and Case-II are presented in the second and third horizontal blocks of Table~7. For Case-I, we obtain $\delta=89.99^\circ$, while for Case-II we obtain $\delta=270.009^\circ$. The global-fit analysis gives a best-fit value of $\delta\simeq274^\circ$ for IO, with an allowed $3\sigma$ range of $201^\circ$--$335^\circ$. Thus, the Case-II prediction is particularly close to the current best-fit value. The physical Majorana phase $\sigma^\prime$ receives only a very small radiative correction, with predicted values of approximately $-0.000075^\circ$ and $0.00008^\circ$ for Case-I and Case-II, respectively. The corresponding predictions for the unphysical phase $\phi_1$ are numerically identical to those of $\delta$, as follows from the analytical expressions derived in Section~3. The predicted values of $J$ and $|m_{ee}|$ are also consistent with the corresponding experimental constraints.\\
 \indent~ The corresponding low-energy predictions for $\tan\beta=30$ and $50$ are presented in the third and fourth vertical blocks of Table~7, respectively. For $\tan\beta=30~(50)$, 
 we obtain $m_1=0.050123$ $(0.049227)~\mathrm{eV}$ and $m_2=0.051194~(0.049265)~\mathrm{eV}$, respectively. In both cases, the sum of the neutrino masses remains below the cosmological upper bound. For $\tan\beta=30$, however, the predicted mass-squared differences are not simultaneously compatible with the corresponding $3\sigma$ ranges. For $\tan\beta=50$, $|\Delta m_{31}^2|=2.42\times10^{-3}~\mathrm{eV}^2$ lies within the allowed range, whereas $\Delta m_{21}^2=3.70\times10^{-5}~\mathrm{eV}^2$ remains outside the corresponding $3\sigma$ range. The mixing-angle predictions for $\tan\beta=30~(50)$ are $\sin^2\theta_{12}=0.34182~(0.34450)$, $\sin^2\theta_{23}=0.49963~(0.49898)$, and $\sin^2\theta_{13}=0.02226~(0.02229)$. The predicted values of $\sin^2\theta_{13}$ and $\sin^2\theta_{23}$ remain compatible with the corresponding experimental constraints. The radiative correction to $\sin^2\theta_{12}$ is comparatively larger, particularly for the larger values of $\tan\beta$. The CP phases, $J$, and $|m_{ee}|$ exhibit the same qualitative behavior as that found for $\tan\beta=10$.\\
 \indent~ From the above analysis, we find that the high-energy input values given in Table~7 lead to low-energy predictions consistent with the observational data only for $\tan\beta=10$. For $\tan\beta=30$ and $50$, the same set of high-energy inputs yields low-energy predictions for all neutrino parameters that are consistent with the observational data, except for the mass-squared differences, which fall outside their respective $3\sigma$ allowed ranges. We therefore adjust the high-energy input mass eigenvalues for $\tan\beta=30$ and $50$ while keeping the remaining input parameters unchanged from those used for $\tan\beta=10$.
 For $\tan\beta=30$, we choose $m_1^{\mu\tau}=0.0558~\mathrm{eV}$ and $m_2^{\mu\tau}=0.0561~\mathrm{eV}$. These inputs give $\Delta m_{21}^2=7.03\times10^{-5}~\mathrm{eV}^2$ and $|\Delta m_{31}^2|=2.49\times10^{-3}~\mathrm{eV}^2$, both of which are consistent with the corresponding $3\sigma$ ranges. For $\tan\beta=50$, the same choice of input masses does not yield a compatible value of $\Delta m_{21}^2$. We therefore choose $m_1^{\mu\tau}=0.0576~\mathrm{eV}$ and $m_2^{\mu\tau}=0.0580~\mathrm{eV}$. The resulting low-energy masses are $m_1=0.050400~\mathrm{eV}$ and $m_2=0.051100~\mathrm{eV}$, giving $\Delta m_{21}^2=7.11\times10^{-5}~\mathrm{eV}^2$ and $|\Delta m_{31}^2|=2.54\times10^{-3}~\mathrm{eV}^2$. Both mass-squared differences are now compatible with the corresponding experimental $3\sigma$ ranges. The modified mass inputs also lead to small changes in the other low-energy neutrino parameters, which remain compatible with the observational constraints. The resulting predictions are presented in Table~8.\\
 \begin{table}[H]
\centering
\begin{tabular}{lc| c|c}
\hline
 & $\tan\beta=10$& $\tan\beta=30$ &$\tan\beta=50$ \\
\hline
Parameter &$1\ \mathrm{TeV}$& $1\ \mathrm{TeV}$  & $1\ \mathrm{TeV}$  \\
 \hline
$m_1^{\mu\tau}$ & 0-0.057433&0-0.057550 & 0-0.059418 \\
$m_2^{\mu\tau}$ &0-0.056670& 0-0.05680 & 0-0.057033 \\
$\sin\theta_{13}^{\mu\tau}$ & 0.142003-0.154365&0.142683-0.155356 & 0.145247-0.157785 \\
$\sin\theta_{12}^{\mu\tau}$ &0.523455-0.587091& 0.526360-0.588545 & 0.522727-0.589273 \\
\hline
\end{tabular}\caption{ Approximate allowed ranges of $m_i^{\mu\tau}$, $\sin\theta_{12}^{\mu\tau}$ and $\sin\theta_{13}^{\mu\tau}$ for $\tan\beta= 10,\ 30$ and $50$  with $\Lambda_s=  1\ TeV$ for the case of IO.}
\end{table} 
 \begin{table}[H]
\centering
\begin{tabular}{lc|c|c|c}
\hline

& High energy value & \multicolumn{2}{c}{Low energy value at $\Lambda_{EW}$} \\
\hline
& &  $\tan\beta=10$& $\tan\beta=30$ & $\tan\beta=50$ \\
\hline
Parameter & $10^{14}\ \mathrm{GeV}$ &$1\ \mathrm{TeV}$& $1\ \mathrm{TeV}$ & $1\ \mathrm{TeV}$ \\
\hline
$m_3~(\mathrm{eV})$&0&0&0&0\\
$m_1\left(eV\right)$ & 0.054 &0.050482& 0.050123 & 0.049227  \\
$m_2\left(eV\right)$ & 0.055 &0.051255& 0.051194 & 0.049265  \\
$\sin^2\theta_{13}$ & 0.0222 &0.02221& 0.02226 & 0.02229 \\
$\sin^2\theta_{23}$ & 0.5 &0.49999 &0.49963 & 0.49898 \\
$\sin^2\theta_{12}$ & 0.3025 & 0.30811& 0.34182 &  0.34450 \\
$\Delta m_{21}^2 (/ 10^{-5})eV^2$&-& 7.86& 10.85& 3.7\\
$|\Delta m_{31}^2|=|m_1^2| (/ 10^{-3})eV^2$&-&2.54& 2.62& 2.42\\
$\Sigma_i m_i \left(eV\right)$ & - &0.101737& 0.101317 & 0.098492 \\
 \hline
Case-I \\ \hline
$\delta\left(/^\circ\right)$ & 90 &89.99& 89.93 & 89.277\\
$\sigma^\prime\left(/^\circ\right)$ &-& -0.000075  & -0.00072 & -0.00203\\
$\phi_1\left(/^\circ\right)$ & 90 & 89.99& 89.93 & 89.277\\
$J$&- & 0.032221&0.032676 & 0.011880 \\
$|\langle m \rangle_{ee}|$&- &0.050510& 0.049466&  0.048332 \\
\hline
Case-II \\ \hline
$\delta\left(/^\circ\right)$ & 270 &270.009& 270.083  & 270.721  \\
$\sigma^\prime\left(/^\circ\right)$&- &0.00008& 0.00072 & 0.00203 \\
$\phi_1\left(/^\circ\right)$ & 270 &270.009& 270.083 & 270.721\\
$J$&- & -0.032221&-0.032676 &- 0.011880 \\
$|\langle m \rangle_{ee}|$&- &0.050510& 0.049466&  0.048332 \\
\hline
\end{tabular}
\caption{High energy values of parameters at $\Lambda_{\mu\tau}=10^{14} \ GeV$ and corresponding low energy values at $m_t$ scale for $\Lambda_s=1 \ TeV$ with $\tan\beta= 10,\ 30$ and $50$  for the case of IO.}
\end{table}

\begin{table}[t]
\centering
\renewcommand{\arraystretch}{1.15}
\begin{tabular}{l|cc|cc}
\hline
& \multicolumn{2}{c|}{$\tan\beta=30$} 
& \multicolumn{2}{c}{$\tan\beta=50$} \\
\hline
Parameter & Input at $\Lambda_{\mu\tau}$ & Output at $\Lambda_{EW}$ & Input at $\Lambda_{\mu\tau}$ & Output at $\Lambda_{EW}$ \\
\hline
$m_3~(\mathrm{eV})$&0&0&0&0\\
$m_1~(\mathrm{eV})$ 
& 0.0558 & 0.049900 & 0.0576 & 0.050400 \\

$m_2~(\mathrm{eV})$ 
& 0.0561 & 0.050600 & 0.0580 & 0.051100 \\

$\sin^2\theta_{13}$ 
& 0.0222 & 0.02226 & 0.0222 & 0.02227 \\

$\sin^2\theta_{23}$ 
& 0.5 & 0.49963 & 0.5 & 0.49998 \\

$\sin^2\theta_{12}$ 
& 0.3025 & 0.32501 & 0.3025 & 0.34322 \\

$\Delta m_{21}^2~(10^{-5}~\mathrm{eV}^2)$ 
& -- & 7.03 & -- & 7.11 \\

$|\Delta m_{31}^2|~(10^{-3}~\mathrm{eV}^2)$ 
& -- & 2.49 & -- & 2.54 \\

$\sum_i m_i~(\mathrm{eV})$ 
& -- & 0.1005 & -- & 0.1015 \\

\hline
\multicolumn{5}{c}{\textbf{Case-I}}\\
\hline

$\delta~(^\circ)$ 
& 90 & 89.83 & 90 & 89.92 \\

$\sigma^\prime~(^\circ)$ 
& -- & -0.0007 & -- & -0.0008 \\

$\phi_1~(^\circ)$ 
& 90 & 89.83 & 90 & 89.92 \\

$J$ 
& -- & 0.033824 & -- & 0.031520 \\

 $|\langle m \rangle_{ee}|$  
& -- & 0.050520 & -- & 0.05148 \\

\hline
\multicolumn{5}{c}{\textbf{Case-II}}\\
\hline

$\delta~(^\circ)$ 
& 270 & 270.083 & 270 & 270.156 \\

$\sigma^\prime~(^\circ)$ 
& -- & 0.00073 & -- & 0.00074 \\

$\phi_1~(^\circ)$ 
& 270 & 270.083 & 270 & 270.156 \\

$J$ 
& -- & -0.033824 & -- & -0.031520 \\

$|\langle m \rangle_{ee}|~(\mathrm{eV})$ 
& -- & 0.050520 & -- & 0.05148 \\

\hline
\end{tabular}
\caption{Low-energy neutrino parameters with high-energy mass inputs newly set for $\tan\beta=30$ and $50$ with $\Lambda_s=1 \ TeV$ for the case of IO.}
\label{tab:adjusted_inputs}
\end{table}
\indent~ We finally estimate the amount of deviations of the low-energy neutrino parameters from their corresponding high-energy input values. The amount of deviations are defined in a similar way as adopted in the NO scenario: $\Delta m_i=m_i^{\mu\tau}-m_i$ for $i=1,2$, $\Delta(\sin^2\theta_{ij})=\sin^2\theta_{ij}^{\mu\tau}-\sin^2\theta_{ij}$, and $\Delta x=x^{\mu\tau}-x$ for the relevant CP phases. The resulting deviations are presented in Table~9. From these data, it is apparent that all the deviations $\Delta m_i$, $\Delta(\sin^2\theta_{ij})$ and $\Delta x$ increase with $\tan\beta$. These trends can be understood from the analytical expressions given in Eqs.~(\ref{eq:77})--(\ref{eq:79}) and (\ref{eq:97})--(\ref{eq:100}), following the same reasoning discussed in the NO scenario.

 \begin{table}[H]
\centering
\begin{tabular}{lccc}
\hline
Parameter & \multicolumn{3}{c}{Amount of deviations at $\Lambda_{EW}$ }\\
\cline{2-4}
& $\tan\beta=10$ & $\tan\beta=30$ & $\tan\beta=50$\\
\hline
$\Delta m_3~(\mathrm{eV})$ &0&0&0\\
$\Delta m_1(\mathrm{eV})$ &0.003518&0.003877&0.004773\\
$\Delta m_2~(\mathrm{eV})$ &0.003745&0.003806&0.005735\\
$\Delta(\sin^2\theta_{13})$ &0.00001&0.00006&0.00009\\
$\Delta(\sin^2\theta_{23})$ &0.00001&0.00037&0.00102\\
$\Delta(\sin^2\theta_{12})$ &0.00561&0.03932&0.042\\
\hline
\multicolumn{4}{c}{\textbf{Case-I}}\\
\hline
$\Delta\delta~(^\circ)$ &0.01&0.07&0.723\\
$\Delta\phi_1~(^\circ)$ &0.01&0.07&0.723\\
\hline
\multicolumn{4}{c}{\textbf{Case-II}}\\
\hline
$\Delta\delta~(^\circ)$ &0.009&0.083&0.724\\
$\Delta\phi_1~(^\circ)$ &0.009&0.083&0.724\\
\hline
\end{tabular}

\caption{Amount of deviations of parameters  at $\Lambda_{EW}$ scale with respect to their high-energy input values for $\Lambda_s=1 \ TeV$ with $\tan\beta= 10,30, 50$ for the case of IO.}
\end{table}

\section{Summary and discussion}
\indent~The recent measurement of the Dirac CP phase, $\delta =274^\circ$ for IO, has strengthened the motivation for studying $\mu$-$\tau$ reflection symmetry as a viable flavor symmetry \cite{T2K1,Nova2,global2}. This symmetry predicts $\theta_{23}=\pi/4$, $\delta=\pi/2$ or $3\pi/2$ and a non-zero value of $\theta_{13}$. In this work, we study the deviations from the $\mu$-$\tau$ reflection symmetry under radiative corrections in the minimal seesaw framework. Starting from the integral solution of the one-loop RGE for the effective Majorana neutrino mass matrix, we obtain the low-energy mass matrix in terms of the high-energy mass matrix. We assume that $\mu$-$\tau$ reflection symmetry is exact at the high-energy scale and take the corresponding reflection-symmetric mass matrix as the high-energy mass matrix at the seesaw scale, $\Lambda_{\mu\tau}=10^{14}$ GeV. We then derive analytical expressions for the low-energy neutrino parameters in terms of their high-energy parameters, considering the NO and IO scenarios separately. We take the top-quark mass scale, $m_t=172.76\ GeV$, as the low-energy scale and use the analytical expressions to estimate the low-energy neutrino parameters from the high-energy inputs. The atmospheric mixing angle and CP phases are fixed by $\mu$-$\tau$ reflection symmetry, while the reactor and solar mixing angles and the neutrino mass eigenvalues are determined from the corresponding correlation plots. For the numerical analysis, we consider the SUSY breaking scale $\Lambda_{\mathrm{SUSY}}=1\ TeV$ and three representative values of $\tan\beta=10$, $30$ and $50$.
 We also perform the analysis separately for the NO and IO scenarios. For the NO scenario, an appropriate choice of high-energy input parameters reproduces all low-energy observables within the current experimental limits for $\tan\beta=10$, $30$ and $50$. The sum of neutrino masses satisfies the cosmological bound, the mass-squared differences remain close to the global best-fit values and the predicted values of $\theta_{12}$ and $\theta_{13}$ agree well with the experimental best-fit values. The predicted $\theta_{23}$ lies within its $3\sigma$ allowed range and shifts toward the second octant due to radiative corrections. We further estimate the amount of deviations of the low-energy parameters from their corresponding high-energy inputs and find that the deviations in the neutrino masses and mixing parameters increase with increasing $\tan\beta$. For the IO case, an appropriate set of high-energy input parameters yields low-energy neutrino parameters consistent with the experimental constraints for $\tan\beta=10$. When the same input parameters are used for $\tan\beta=30,\ 50$ and all observables remain consistent with the experimental constraints except the mass-squared differences, which fall outside their respective $3\sigma$ ranges. By choosing suitable high-energy input parameters for $\tan\beta=30$ and $50$, we recover mass-squared differences within the allowed ranges, while the resulting low-energy neutrino parameters remain consistent with the experimental data
 A notable feature in the IO case is that the low-energy value of $\theta_{23}$ prefers the first octant under radiative corrections. This is opposite to the behavior observed in the NO scenario. Furthermore, the predicted Dirac CP phase, $\delta\simeq270.721^\circ$ for Case-II, is particularly noteworthy because it lies very close to the experimentally measured value of $\delta=274^\circ$. The overall pattern of deviations of the low-energy neutrino parameters from their high-energy inputs in the IO scenario is qualitatively similar to that found for NO.


\begin{thebibliography}{90}
\bibitem{SK2} K. Abe {\it et al.}, (SK Collab.), {\it Phys. Rev. Lett.} {\bf 134}, 12488 (2025)
\bibitem{SNO1} A. Allega {\it et al.}, (SNO+ Collab.), {\it Eur. Phys. J. C} {\bf 85}, 17 (2025)
\bibitem{MINOS1} P. Adamson {\it et al.}, (MINOS+ Collab.), {\it Phys. Rev. Lett.} {\bf 125}, 131802 (2020)
\bibitem{DAYA1} F. P. An {\it et al.}, (Daya Bay Collab.), {\it Phys. Rev. Lett.} {\bf 135}, 201802 (2025)
\bibitem{Reno} G. Bak {\it et al.}, (RENO Collab.), {\it Phys. Rev. Lett.} {\bf 121}, 201801 (2018)
\bibitem{Dchooz} H. de Kerret {\it et al.}, (Double Chooz Collab.), {\it Nat. Phys.} {\bf 16}, 558 (2020)
\bibitem{T2K1} K. Abe {\it et al.}, (T2K Collab.), {\it Phys. Rev. D} {\bf 108}, 072011 (2023)
\bibitem{Nova2} J. Wolcott, (NOvA Collab.), {\it DOI:10.2172/2429313},(2024)
\bibitem{ICE3} R. Abbasi {\it et al.}, (IceCube Collab.), {\it Phys. Rev. Lett.} {\bf 134}, 091801 (2025)
\bibitem{global2} I. Esteban {\it et al.}, {\it JHEP} {\bf 2024}, 1 (2025)\bibitem{Mass1} SR. Choudhury and S. Hannestad, {\it JCAP} {\bf 2020}, 37 (2020)
\bibitem{Kam} S. Abe{\it et al.}, {\it arXiv:2406.11438v1 [hep-ex]}  (2024).
\bibitem{ref1} P. F. Harrison, W. G. Scott, {\it Phys. Lett. B} {\bf 547}, 219–228 (2002)
\bibitem{rev1} Z-z. Xing and Z-h. Zhao, {\it Rep.  Prog. Phys} {\bf 79}, 076201 (2016)
\bibitem{vien1} V. V. Vien, {\it Chin. J. Phys.} {\bf 93}, 418--429 (2025)
\bibitem{vien2} V. V. Vien, {\it Phys. Lett. B} {\bf 859}, 139132 (2024)
\bibitem{vien3} V. V. Vien, {\it Phys. Lett. B} {\bf 858}, 139061 (2024)
\bibitem{king5} S. F. King and Y-L. Zhou, {\it JHEP} {\bf 05}, 217 (2019)
\bibitem{zhao} Z-h. Zhao {\it Eur. Phys. J. C} {\bf 82}, 436 (2022)
\bibitem{xinggg} Z-z. Xing and J-y. Zhu, {\it Chin. Phys. C} {\bf 41}, 123103 (2017)
\bibitem{nnath1} N. Nath, {\it Phys. Rev. D} {\bf 99}, 035026 (2019)
\bibitem{nishi} C. C. Nishi and  B.L. S´anchez-Vega, {\it JHEP} {\bf 01}, 068 (2017)
\bibitem{nishi1} C. C. Nishi, {\it Phys. Rev. D} {\bf 93}, 093009 (2016)
\bibitem{dirac1} Z-z. Xing, D. Zhang and J-y. Zhu, {\it JHEP} {\bf 11}, 135 (2017)
\bibitem{zhou} Y-L. Zhou, arXiv:1409.8600v1 [hep-ph]
\bibitem{original} P.H. Frampton, S.L.Glashow and T. Yanagida, {\it Phys. Lett. B} {\bf 548}, 119-121 (2002)
\bibitem{mini1} E. Ma and D.P. Roy, {\it Phys. Rev. D} {\bf 46}, 3000-3007 (1992)
\bibitem{mini2} E. Ma and D.P. Roy, {\it Phys. Rev. D} {\bf 59}, 097702 (1999)
\bibitem{mini3} T. Kitabayashi, {\it Phys. Rev. D} {\bf 76}, 033002(2007)
\bibitem{mini4} X-G. He {\it et al.}, {\it Phys. Rev. D} {\bf 91}, 076008(2015)
\bibitem{prog} Z-z Xing and Z-h Zhao, {\it Rep. Prog. Phys.} {\bf 84}, 066201(2021)
\bibitem{Ssy13} E. Gibney, {\it Nature} {\bf 605}, 604-607 (2022)
\bibitem{Ssy14} R. Aaij {\it et al.} (LHCb), {\it Nature} {\bf 11}, 743 (2015)
\bibitem{Ssy15} M. Kazana (CMS),  {\it Acta Phys. Polon} {\bf 47}, 1489 (2016)
\bibitem{nnath} N. Nath,  Z-z. Xing and J. Zhang, {\it Eur. Phys. J. C} {\bf 78}, 289 (2018)
\bibitem{liu} Z-C. Liu, C-X. Yue and Z-h. Zhao, {\it JHEP} {\bf 10}, 1-25 (2017)
\bibitem{duarah} C. Duarah, {\it Phys. Lett. B} {\bf 815}, 136119 (2021)
\bibitem{chank} P. H. Chankowski and Z. Pluciennik, {\it Phys. Lett. B} {\bf 316}, 312 (1993)
\bibitem{babu} K. S. Babu, C. N. Leung and J Pantaleone, {\it Phys. Lett. B} {\bf 319}, 191 (1993)
\bibitem{casas} J. A. Casas, J. R. Espinosa, A. Ibarra and I. Navarro {\it Nucl. Phys. B} {\bf 573}, 652 (2000)
\bibitem{antusch} S. Antusch {\it et al.}, {\it Phys. Lett. B} {\bf 519}, 238 (2001)
\bibitem{antusch1} S. Antusch, J. Kersten and M. Lindner and M. Ratz, {\it Nucl. Phys. B} {\bf 674}, 401 (2003)
\bibitem{chank1} P. H. Chankowski and S. Pokorski, {\it Int. J. Mod. Phys. A} {\bf 17}, 575 (2002)
\bibitem{ray} S. Ray, {\it Int. J. Mod. Phys. A} {\bf 25}, 4339 (2010)
\bibitem{ingr1} J. Ellis and S. Lola, {\it Phys. Lett. B} {\bf 458}, 310-321 (1999)
\bibitem{ingr2} P. H. Chankowski, W. Krolikowski and S. Pokorski, {\it Phys. Lett. B} {\bf 473}, 109-117 (2000)
\bibitem{pegu} P. Pegu and C. Duarah, {\it Eur. Phys. J. C} {\bf 85}, 959 (2025)
\end{thebibliography}
\end{document}